\documentclass[cernpreprint,USenglish,a4paper]{na61doc}
\usepackage{amsmath}
\usepackage{lineno}
\usepackage{multirow}
\usepackage{makecell}
\usepackage{comment}
\usepackage[normalem]{ulem}

\usepackage{colortbl}
\definecolor{darkred}{rgb}{0.5,0,0}
\definecolor{darkblue}{rgb}{0,0,0.5}
\definecolor{firebrick}{rgb}{0.75,0.125,0.125}
\definecolor{darkgreen}{rgb}{0,0.5,0}

\usepackage[colorlinks=true,linkcolor=firebrick,citecolor=darkgreen,urlcolor=darkblue]{hyperref}
\usepackage{color}
\usepackage{xspace}
\usepackage{enumerate}
\DeclareUnicodeCharacter{2212}{-}
\usepackage{cite}

\ShineTitle{$K^{0}_{S}$ meson production in inelastic \textit{p+p} interactions at 158~\GeVc beam momentum measured by \NASixtyOne at the CERN SPS}

\PreprintIdNumber{CERN-EP-2021-112}

\ShineJournal{}	

\ShineJournalRef{}
\ShineDOI{}

\ShineAbstract{
The production of $K^{0}_{S}$ mesons in inelastic \textit{p+p} collisions at beam momentum 158~\GeVc ($\sqrt{s_{NN}}=17.3$~\GeV) was measured with the \NASixtyOne spectrometer at the CERN Super Proton Synchrotron. Double-differential distributions were obtained in transverse momentum and rapidity. The mean multiplicity of $K^{0}_{S}$ was determined to be $0.162 \pm 0.001 (stat.) \pm 0.011 (sys.)$. The results on $K^{0}_{S}$ production are compared with model predictions (\EposLong, SMASH 2.0, PHSD and UrQMD 3.4 models) as well as with published world data. 
}

\newcommand{\eV}{\ensuremath{\mbox{e\kern-0.1em V}}\xspace}
\newcommand{\GeV}{\ensuremath{\mbox{Ge\kern-0.1em V}}\xspace}
\newcommand{\MeV}{\ensuremath{\mbox{Me\kern-0.1em V}}\xspace}
\newcommand{\GeVc}{\ensuremath{\mbox{Ge\kern-0.1em V}\!/\!c}\xspace}
\newcommand{\GeVcc}{\ensuremath{\mbox{Ge\kern-0.1em V}\!/\!c^2}\xspace}
\newcommand{\MeVcc}{\ensuremath{\mbox{Me\kern-0.1em V}\!/\!c^2}\xspace}
\newcommand{\AGeV}{\ensuremath{A\,\mbox{Ge\kern-0.1em V}}\xspace}
\newcommand{\AGeVc}{\ensuremath{A\,\mbox{Ge\kern-0.1em V}\!/\!c}\xspace}
\newcommand{\MeVc}{\ensuremath{\mbox{Me\kern-0.1em V}/c}\xspace}

\newcommand{\cm}{\ensuremath{\mbox{cm}}\xspace}

\newcommand{\dd}{\ensuremath{{\text{d}}}\xspace}
\newcommand{\dedx}{\ensuremath{\dd E\!/\!\dd x}\xspace}

\newcommand{\pt}{\ensuremath{p_{\text{T}}}\xspace}

\newcommand{\coordinate}[1]{{\fontfamily{lmss}\selectfont#1}}

\newcommand{\GeantThree}{{\scshape Geant3}\xspace}

\newcommand{\EposLong}{{\scshape EPOS 1.99}\xspace}

\newcommand{\CernVM}{\textsc{Cern\-\kern-0.05emVM}\xspace}

\begin{document}

\maketitle


\section{Introduction}\label{sec:intro}

Kaons contain strange or anti-strange valence quarks, which are both not present in the initial state of collisions between nucleons and nuclei. Thus kaon production implies the creation of a strange and anti-strange quark pair. Collisions between nuclei proceed via the formation of a rapidly expanding high energy density fireball~\cite{Florkowski:2010zz}. 
At sufficiently high collision energy, the evolution of the fireball is expected to proceed via an intermediate partonic phase, the quark-gluon plasma (QGP). Thus, investigating these reactions will shed light on the differences between hadronic and partonic matter and the characteristics of the phase transition between them. The study of kaon production in \textit{p+p} collisions is important not only as a reference for possible modifications of strangeness production in nucleus-nucleus collisions~\cite{Rafelski:1982pu} but also for understanding strangeness production in elementary interactions. It was predicted that the onset of deconfinement is located in the few \GeV energy range \cite{Gazdzicki:1998vd}. In order to explore this region systematically \NASixtyOne studies observables indicative of the QGP by a two-dimensional scan in collision energy and nuclear mass number of the colliding nuclei. Since 2009 \NASixtyOne has collected data on \textit{p+p}, p+Pb, Pb+Pb, Be+Be, Ar+Sc and Xe+La interactions in the energy range 13A-158A \GeV \cite{Gazdzicki:995681}. Results on identified hadron spectra measurements can be found in Ref. \cite{NA61SHINE:2017fne, NA61SHINE:2020czq ,NA61SHINE:2021nye, NA61SHINE:2020dwg, NA61SHINE:2015haq}. In this paper we present the first results of $K_{S}^{0}$ production in \textit{p+p} collisions at 158 \GeV, which will be used later as the reference for comparison with $K_{S}^{0}$ production at lower energies and constitutes the first step of the energy scan of $K_{S}^{0}$ production in \textit{p+p} interactions. After the energy scan we will perform the nuclear mass scan (heavier systems) and results will be compared with \textit{p+p} collisions. Thanks to high statistics, large acceptance and good resolution the results presented here have significantly higher precision than previously published data at the SPS energies \cite{Ammosov1976, Brick:1980vj, AstonGarnjost:1975im, Chapman:1973fn, Jaeger1974pk, Sheng1976, Lopinto:1980ct, Bailly1987}.

The paper is organised as follows. In section~\ref{sec:setup}, details of the \NASixtyOne detector system are presented. Section \ref{sec:method} is devoted to the description of the analysis method. The results are shown in section~\ref{sec:results}. In section~\ref{sec:comparison}, they are compared to published world data and model calculations. Section~\ref{sec:summary} closes the paper with a summary and outlook.

The following variables and definitions are used in this paper. The particle rapidity $y$ is calculated in the proton-proton collision center of mass system (cms), $y=0.5ln[(E+cp_L)/(E-cp_L)]$, where $E$ and $p_L$ are the particle energy and longitudinal momentum, respectively. The transverse component of the momentum is denoted as $p_T$. The momentum in the laboratory frame is denoted $p_{lab}$ and the collision energy per nucleon pair in the centre of mass by $\sqrt{s_{NN}}$.

\section{Experimental setup}\label{sec:setup}

\begin{figure*}[ht]
  \centering
  \includegraphics[width=0.8\textwidth]{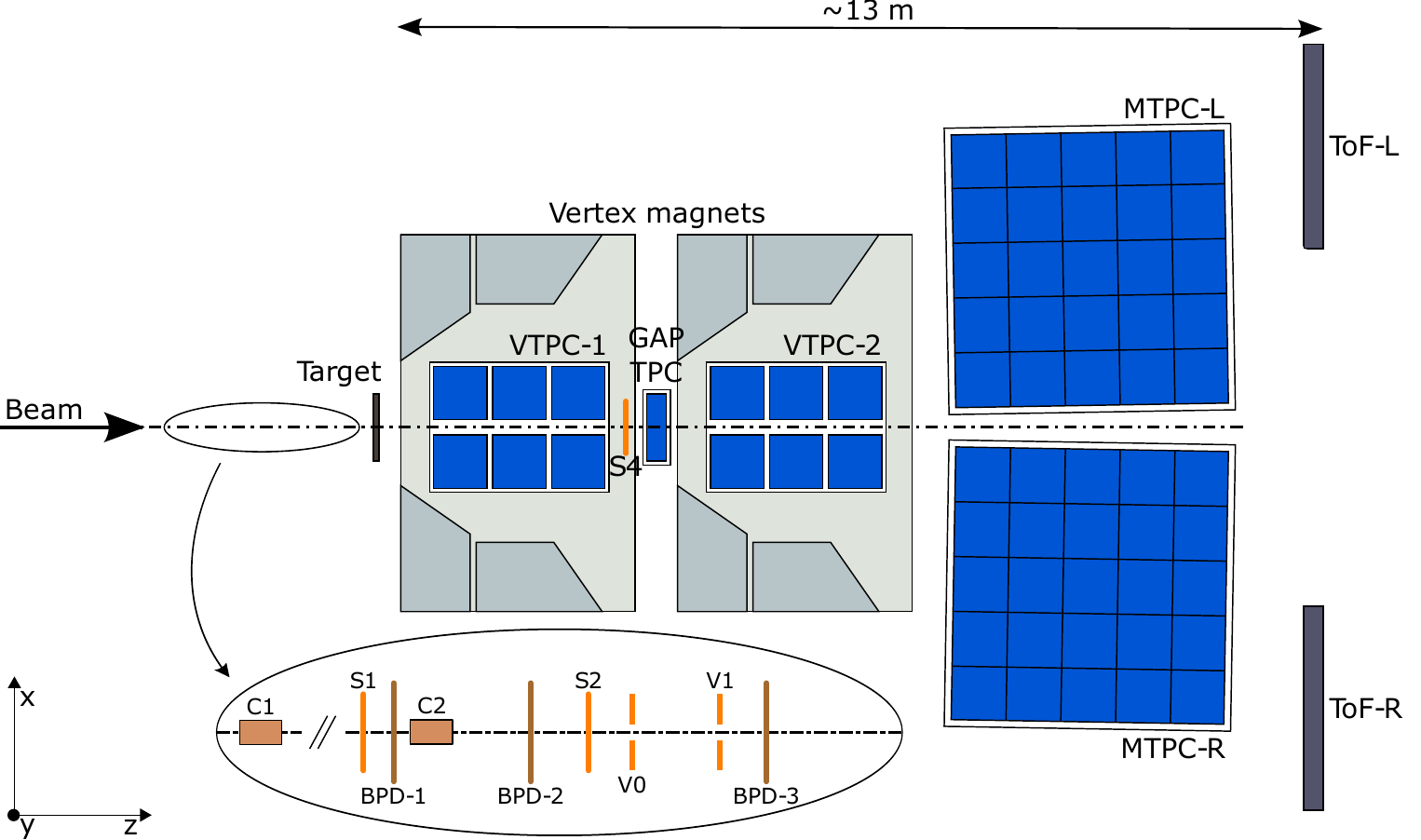}
  \caption[]{
    (Color online) The schematic layout of the NA61/SHINE experiment at the CERN SPS during \textit{p+p} 158 \GeVc data taking (horizontal cut, not to scale). The beam and trigger detector configuration used for data taking in 2009 is shown in the inset (see Refs.~\cite{NA61SHINE:2013tiv, Aduszkiewicz:2015jna} for detailed description). The chosen coordinate system is drawn on the lower left: its origin lies in the middle of the VTPC-2, on the beam axis. }
  \label{fig:detector-setup}
\end{figure*}

The \NASixtyOne collaboration uses a large acceptance spectrometer located in the CERN North Area. The schematic layout of the \NASixtyOne detector during the \textit{p+p} 158 \GeVc data-taking is shown in Fig.~\ref{fig:detector-setup}. A detailed description of the full detector can be found in Ref.~\cite{Abgrall:2014xwa}, while the details on the performance of the simulation in describing the detector performance across different kinematic variables as well as its inefficiencies can be found in Ref. \cite{NA61SHINE:2013tiv}.

The main components of the spectrometer used in this analysis are four large volume Time Projection Chambers (TPC). Two of them, the vertex TPCs (\mbox{VTPC-1} and \mbox{VTPC-2}), are located in the magnetic fields of two super-conducting dipole magnets with a maximum combined bending power of 9~Tm which corresponds to about 1.5~T and 1.1~T fields in the upstream and downstream magnets, respectively. Two large main TPCs (\mbox{MTPC-L} and \mbox{MTPC-R}) and two walls of pixel Time-of-Flight (ToF-L/R) detectors are positioned symmetrically to the beamline downstream of the magnets. A GAP-TPC (GTPC) is placed between \mbox{VTPC-1} and \mbox{VTPC-2} directly on the beamline. It closes the gap between the beam axis and the sensitive volumes of the other TPCs. The TPCs are filled with Ar and CO$_2$ gas mixtures. Particle identification in the TPCs is based on measurements of the specific energy loss (\dedx) in the chamber gas. 

Secondary beams of positively charged hadrons at 158~\GeVc are produced from 400~\GeVc proton beams extracted from the SPS accelerator. Particles of the secondary hadron beam are identified by two Cherenkov counters, a CEDAR-N~\cite{Bovet:1982xf} and a threshold counter (THC). The CEDAR counter, using a coincidence of six out of the eight photomultipliers placed radially along the Cherenkov ring, provides identification of protons, while the THC, operated at a pressure lower than the proton threshold, is used in anti-coincidence in the trigger logic. A selection based on signals from the Cherenkov counters allowed one to identify beam protons with a purity of about 99\%. 
A set of scintillation (S1 and S2), veto (V0 and V1) and Cherenkov counters (C1 and C2) and beam position detectors (BPDs) upstream of the spectrometer provide timing reference, identification, and position measurements of incoming beam particles. The trigger scintillation counter S4 placed downstream of the target has a diameter of 2 cm and is used to select events with collisions in the target area by the absence of a charged particle hit. 

A cylindrical target vessel of 20.29 cm length and 3 cm diameter was situated upstream of the entrance window of VTPC-1 (centre of the target \coordinate{z} = -581~\cm in the NA61/SHINE coordinate system). The vessel was filled with liquid hydrogen corresponding to an interaction length of 2.8\%. The ensemble of the vessel and liquid hydrogen constitute the "Liquid Hydrogen Target" (LHT). Data were taken with full and empty LHT.

Interactions in the target are selected with the trigger system by requiring an incoming beam proton and no signal from the S4 counter. This minimum bias trigger is based on the disappearance of the beam proton downstream of the target.


\section{Analysis}\label{sec:method}

\subsection{Data set}
\begin{figure*}[hb]
  \centering
\includegraphics[width=0.70\textwidth]{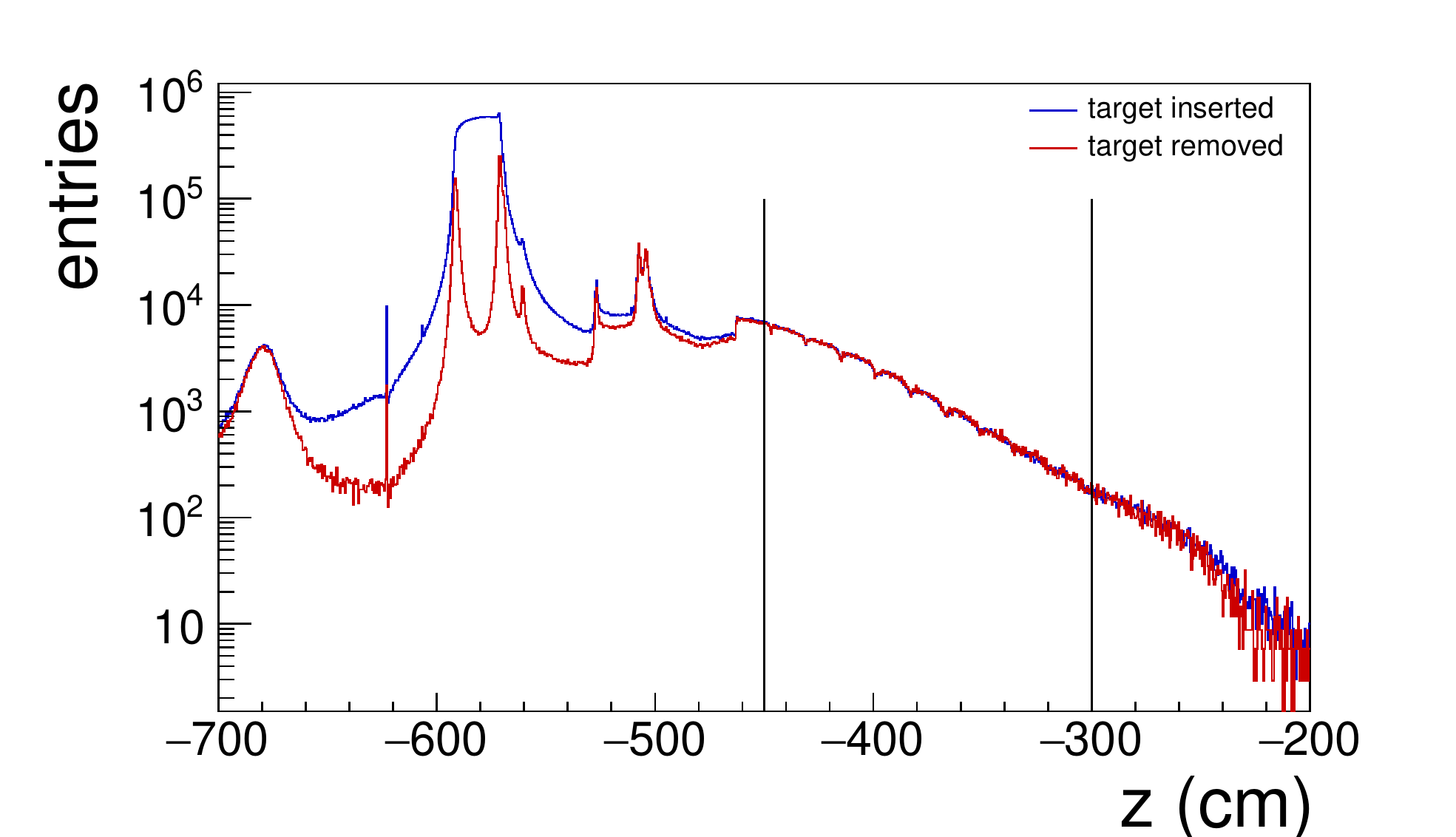}
\vspace{0.0cm}
 \caption[]{Distributions of the \coordinate{z}-coordinate of the reconstructed vertex for events recorded with target full (blue histogram) and target empty (red histogram). Two vertical black lines at position -450 and -300~\cm show the range which is used for histogram normalisation. 
 }
  \label{fig:vertexz}
\end{figure*}

In 2009, 2010 and 2011 the \NASixtyOne detector registered about $5.75 \times 10^7$ \textit{p+p} interactions at 158~\GeVc. For the analysis, the range of the \coordinate{z}-position of the main vertex was selected to cover mostly the LHT (see Sec.~\ref{s:event_selection}) in order to maximize the number of good events and minimize the contamination by off-target interactions. Figure~\ref{fig:vertexz} shows the distributions of reconstructed vertex \coordinate{z} positions in the target-inserted and the target-removed sample as blue and red histograms, respectively. The target-removed sample was normalised in the range -450 < \coordinate{z} < -300~\cm to the same number of reconstructed events as in the target-inserted sample. The normalised ratio of events in the range -590 < \coordinate{z} < -572~\cm is small at the level of 2.9\%, and therefore no correction for non-target interactions was applied. In order to estimate the possible systematic biases related to the contamination by off-target interactions, the event selection window of the \coordinate{z}-position of the main vertex was varied (see Sec.~\ref{s:systematic_uncertainties}). 

\subsection{Analysis method}
\label{sec:analysis_method}


Details of the track and vertex reconstruction procedures can be found in Refs.~\cite{NA61SHINE:2013tiv, Aduszkiewicz:2015jna, Aduszkiewicz:2016mww}. In the following section, the criteria for the selection of events, of tracks and of the $K^{0}_{S}$ decay topology are enumerated. Then the simulation-based procedure will be described, which is used to quantify the losses due to reconstruction inefficiencies and the limited geometrical acceptance.

\subsection{Event selection}
\label{s:event_selection}

The selection criteria for inelastic \textit{p+p} interactions are the following:

\begin{itemize}
	\item [(i)] An interaction was accepted by the trigger logic (see Refs.~\cite{NA61SHINE:2013tiv, Aduszkiewicz:2015jna}).
	\item [(ii)] Beam particle trajectory measured in at least three planes out of four of BPD-1 and BPD-2 and in both planes of BPD-3.
	\item [(iii)] The primary interaction vertex fit converged.
	\item [(iv)] \coordinate{Z} position of the interaction vertex (fitted using the beam trajectory and TPC tracks) not farther away than 9~\cm from the center of the Liquid Hydrogen Target.
\end{itemize}

The final number of events that satisfy all the above selection criteria is $2.86 \times 10^7$.

\subsection{Track and topology selection}
\label{s:track_selection}

\begin{figure*}[ht!]
\vspace{-1.5cm}
\centering
\includegraphics[width=0.7\textwidth]{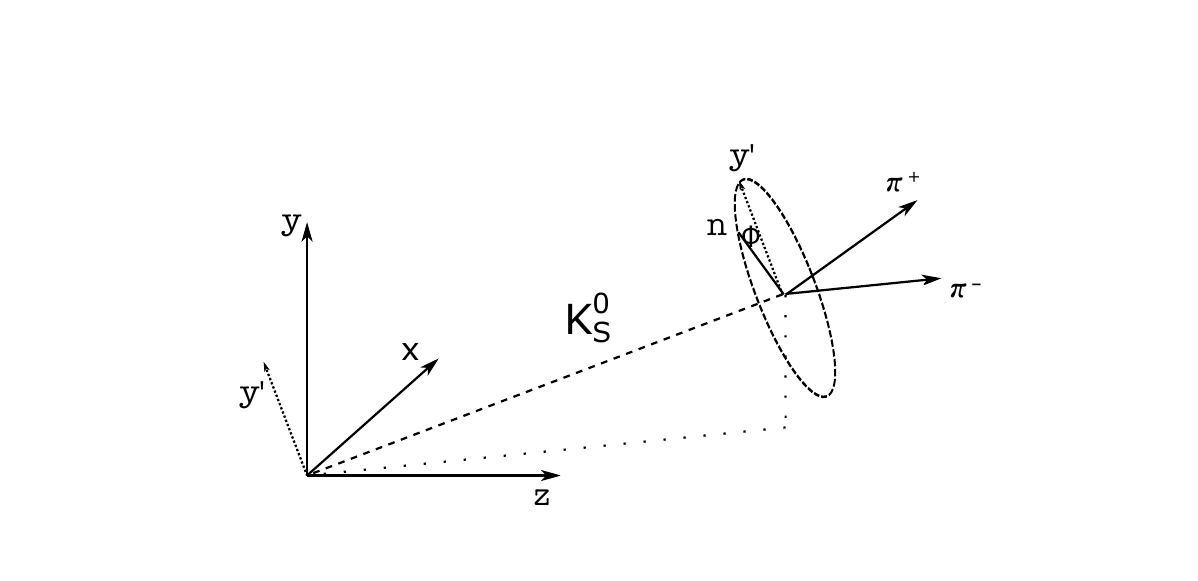}
 \caption[]{Definition of angle $\Phi$ used for $V^0$ selection. $\Phi$ is defined as the angle between the vectors y', and n, where y' is the vector perpendicular to the momentum of the $V^0$-particle which lies in the plane spanned by the \coordinate{y}-axis and the $V^0$-momentum vector, and n is a vector normal to the decay plane. } 
  \label{fig:Phi}
\end{figure*}

Neutral strange particles are detected and measured by means of their weak decays into a pair of charged particles. The $K^{0}_{S}$ decays into $\pi^+ + \pi^-$ with a branching ratio of 69.2\% ~\cite{PDG}. The decay particles form the so-called $V^0$ topology. $K^{0}_{S}$ decay candidates ($V^0$s) are obtained by pairing all positively and negatively charged pions. The corresponding tracks are required to have a distance of closest approach between the two trajectories of less than 1~\cm. The tracks of the decay pions and the $V^0$ topology are subject to the following additional selection criteria:
\begin{itemize}
	\item [(i)] For each track, the minimum number of measured clusters in VTPC-1 and VTPC-2 was required to be 15.
	\item [(ii)] All pion tracks must have a measured specific energy loss (\dedx) in the TPCs within $\pm 3\sigma$ around the nominal Bethe-Bloch value for charged pions. Here $\sigma$ represents the typical standard deviation of a Gaussian fitted to the \dedx distribution of pions. Since only small variations of $\sigma$ were observed for different bins and beam momenta, a constant value $\sigma = 0.052$ is used \cite{NA61SHINE:2020czr}. This selection criteria is applied only for experimental data, not for MC simulated data (see below).
	\item [(iii)] The orientation of the $V^{0}$ decay plane with respect to the magnetic field, quantified by $|cos\Phi|$ (see Fig.~\ref{fig:Phi}), is required to lie in the ranges $|cos\Phi|<0.95$ for $-0.25<y<0.25$, $|cos\Phi|<0.9$ for $0.25<y<0.75$, $|cos\Phi|<0.8$ for $0.75<y<1.25$ and $|cos\Phi|<0.5$ for larger $y$ ($y$ is the ($K^{0}_S$) rapidity). This criterion removes $V^0$s 
	for which the determination of the momenta of the decay products and the decay vertex position suffer from large uncertainties. 
	\item [(iv)] The distance |$\Delta$\coordinate{z}| between the primary production vertex and the $K^0_S$ decay vertex is required to lie in the rapidity dependent range |$\Delta$\coordinate{z}| > $e^{3.1+0.42 \cdot y}$.
	\item [(v)] Spurious $K^0_S$ candidates are rejected by an elliptic cut on the impact parameters of the daughter tracks, which are relative to the $K^{0}_{S}$ decay vertex, in \coordinate{x} ($b_{x}$) and \coordinate{y} ($b_{y}$) direction, $\left(\frac{b_x}{2}\right)^2 + {b_y}^2 < 0.25$~\cm.
\end{itemize}

\begin{figure*}[ht]
  \centering
  \includegraphics[width=0.49\textwidth]{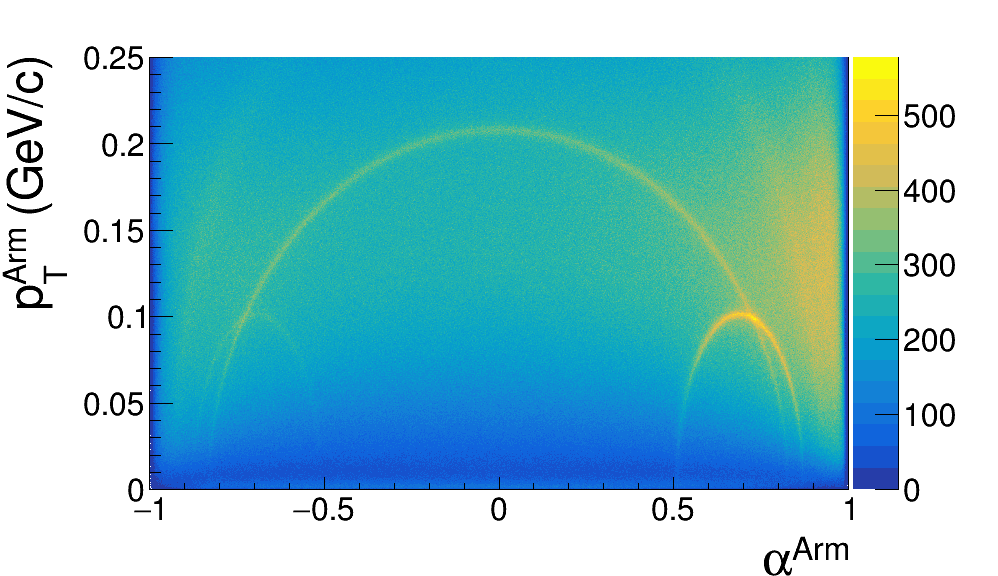}
  \includegraphics[width=0.49\textwidth]{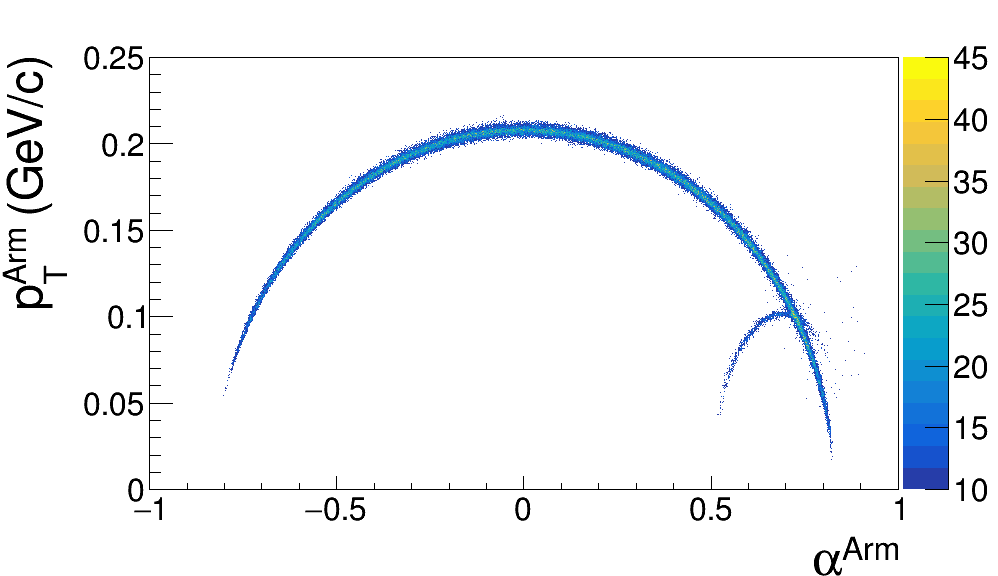} 
  \caption[]{Armenteros-Podolanski plots for $V^0$ candidates without 
  (\textit{left}) and  with all track and topology selection criteria (\textit{right}). 
  }
  \label{fig:APplots} 
\end{figure*}

The quality of the aforementioned track and topology selection criteria is illustrated in Fig.~\ref{fig:APplots}. The population of $K^0_S$ decay candidates is shown as a function of the two Armenteros-Podolansky variables $p^{Arm}_{T}$ and $\alpha^{Arm}$~\cite{Armenteros-Podolanski} before (\textit{left}) and after (\textit{right}) all track and topology selection criteria. The quantity $p^{Arm}_{T}$ is the transverse momentum of the decay particles with respect to the direction of motion of the $V^0$ candidate and $\alpha^{Arm} = (p^{+}_{L} - p^{−}_{L})/(p^{+}_{L} + p^{-}_{L})$, where $p^{+}_{L}$ and $p^{-}_{L}$ are the longitudinal momenta of the positively and negatively charged $V^0$ daughter particles, measured with respect to the $V^0$'s direction of motion. After applying all cuts, a contamination by $\Lambda$'s of roughly 7\% persists (Fig.~\ref{fig:APplots}~\textit{right}). However, the $\Lambda$ background below the $K^0_S$ mass peak is small and amounts to 0.5\%.

\subsection{\texorpdfstring{$K^0_S$}{Lg} yields}
\label{s:signal_extraction}

\begin{figure*}[h]
  \centering
  \includegraphics[width=0.49\textwidth]{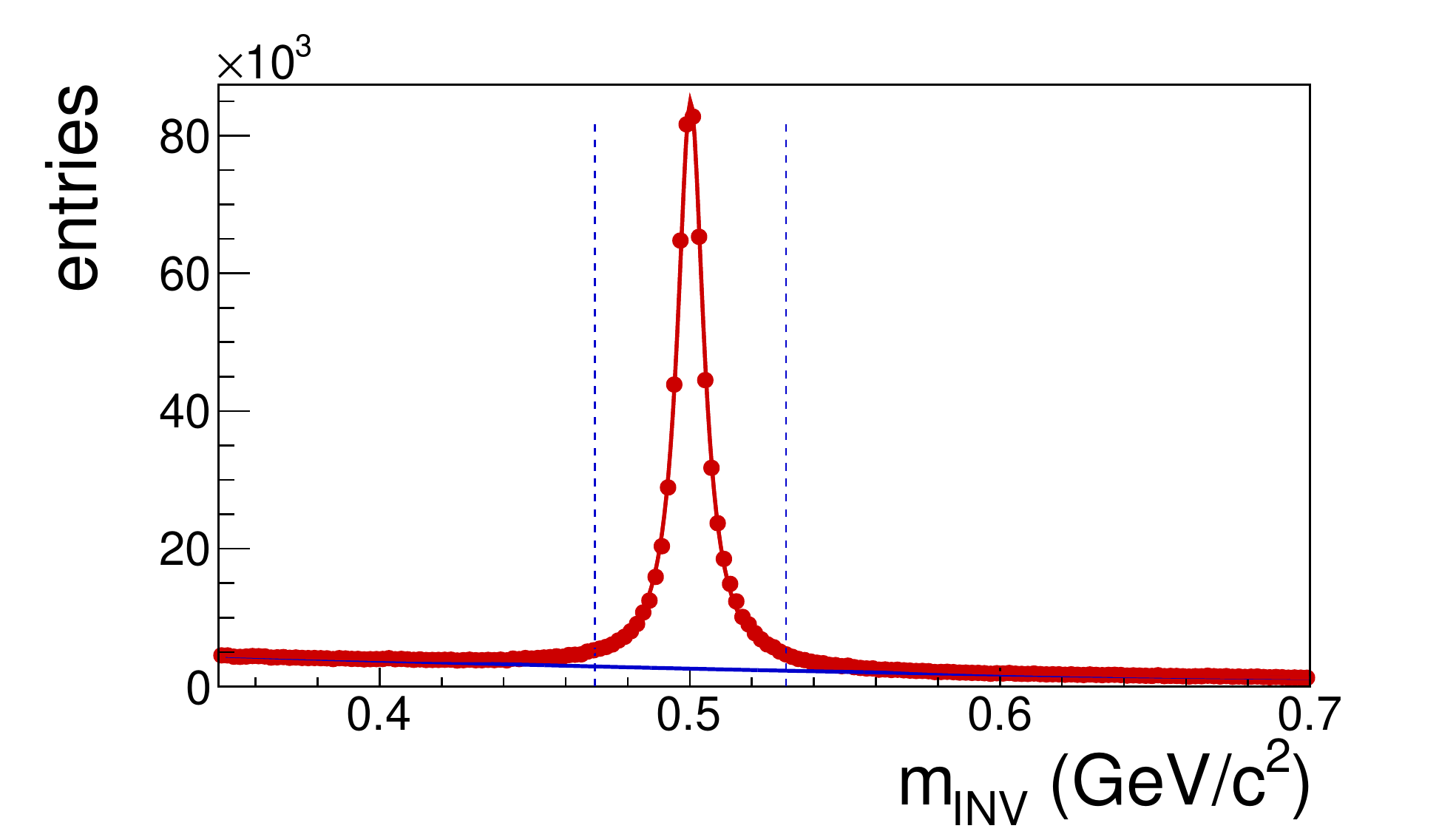}
  \includegraphics[width=0.49\textwidth]{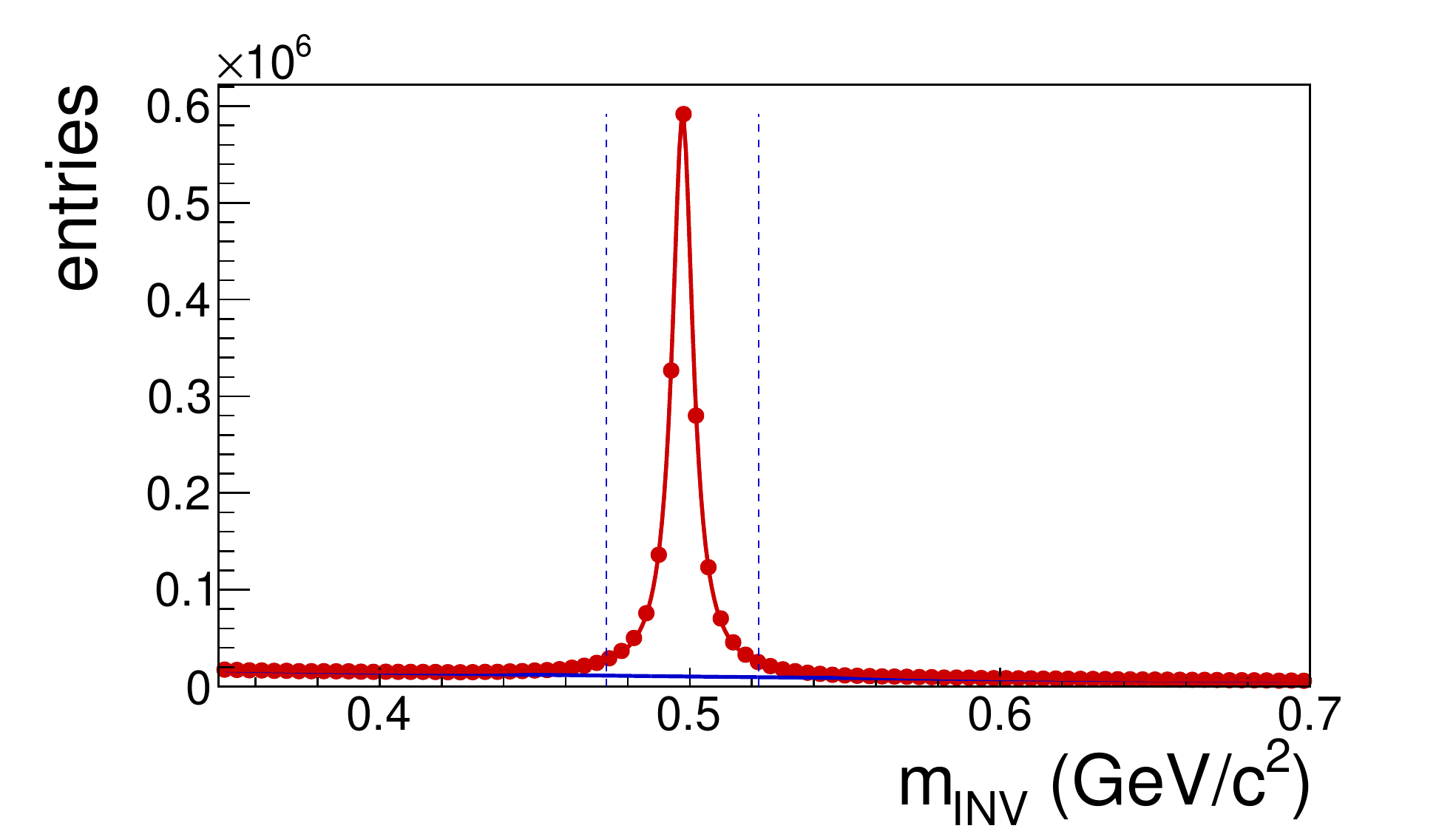} 
  \caption[]{The invariant mass distribution of $K^{0}_{S}$ candidates for experimental data (\textit{left}) and MC (\textit{right}). The dashed vertical lines indicate the regions over which the $K^{0}_{S}$ signal was integrated. The signal data points are shown in red, the fitted background in blue and the total fit results in red. The uncertainties are smaller than symbol size and not visible on the plots. Mass resolutions obtained from the fits are: $\sigma = (0.01026 \pm 0.00002)~\GeVcc$ for the experimental data and $\sigma = (0.00819 \pm 0.00001)~\GeVcc$ for the MC.
  }
  \label{fig:Fits}
\end{figure*}

The double differential yield of $K^0_S$ was determined by studying the invariant mass distributions of the accepted pion pairs in bins of rapidity and transverse momentum (examples are presented in Fig.~\ref{fig:Fits}). True decays will appear as a peak over a smooth background. The $K^0_S$ yield was determined in each bin using a fit function that describes both the signal and the background. A Lorentzian function was used for the signal:

\begin{equation}
L(m) = A \frac{1}{\pi} \frac{\frac{1}{2}\Gamma}{(m-m_0)^2 + (\frac{1}{2}\Gamma)^2}~,
\end{equation}

where A is the normalization factor, $\Gamma$ is the full width at the half maximum of the signal peak, and $m_0$ is the mass parameter. The background contribution is described by a polynomial function of $2^{nd}$ order. Figure~\ref{fig:Fits} shows examples of $\pi^{+} \pi^{-}$ mass distributions after all $V^0$ selection cuts as the red histograms for real events (\textit{left}) and for simulated events (\textit{right}). Clearly, the background outside the $K^{0}_{S}$ peak is small. The width and mass of the $K^0_S$ peak are well reproduced by the simulation, thus indicating that the correction factors used for calculating the final bin-by-bin $K^0_S$ multiplicities are reliable.

The procedure of fitting the histograms proceeds in three steps. In the first step, the background outside the signal peak ([0.475-0.525]~\GeVcc) is fitted with a polynomial of $2^{nd}$ order. This step is necessary to obtain starting values for the parameters of the background function. In the next step, a fit of the full invariant mass spectrum is performed with the sum of the Lorentzian and the background functions. The initial parameter values for the background function are taken from the previous step, while the mass parameter is fixed to the PDG value of $m_0 = 0.497614(24)$ \GeVcc~\cite{PDG} and the width was allowed to vary between 0.01 and 0.03~\GeVcc. Finally, in the last step, all parameters were free, and the fitting region was [0.35-0.7]~\GeVcc. The fitted polynomial background function is shown by the blue curve, and the fitted Lorentzian signal function by the red curve in Fig.~\ref{fig:Fits}. In order to minimize the sensitivity of the $K^0_S$ yield to the integration window, the uncorrected number of $K^{0}_{S}$ was calculated by subtracting bin-by-bin the fitted background and summing the background-subtracted signal in the mass window $m_0 \pm 3\Gamma$ (dashed vertical lines), where $m_0$ is the fitted mass of the $K^{0}_{S}$.
The latter agrees with the PDG value within statistical uncertainties. 
Figure~\ref{fig:Fits} shows that the simulation reproduces the central value of the $K^0_S$ mass distribution and somewhat underestimate its width. To calculate the signal from the simulation,
the $\Gamma$ parameter fitted to the simulation was used. 
Thus a possible bias due to differences between the data and simulation is reduced, 
see Sec.~\ref{s:systematic_uncertainties}.

\begin{figure*}[t]
  \centering
  \includegraphics[width=0.49\textwidth]{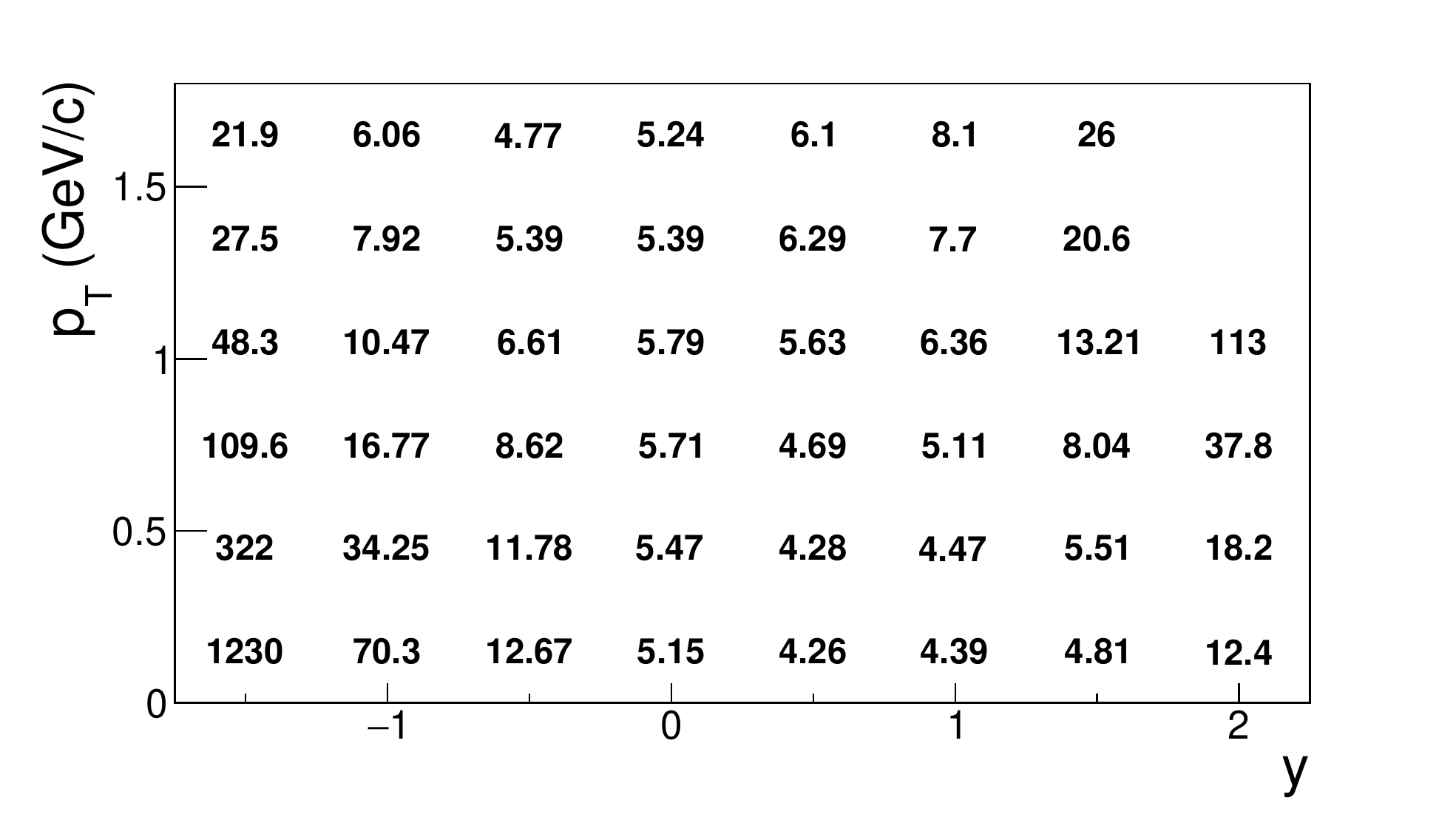}
  \includegraphics[width=0.49\textwidth]{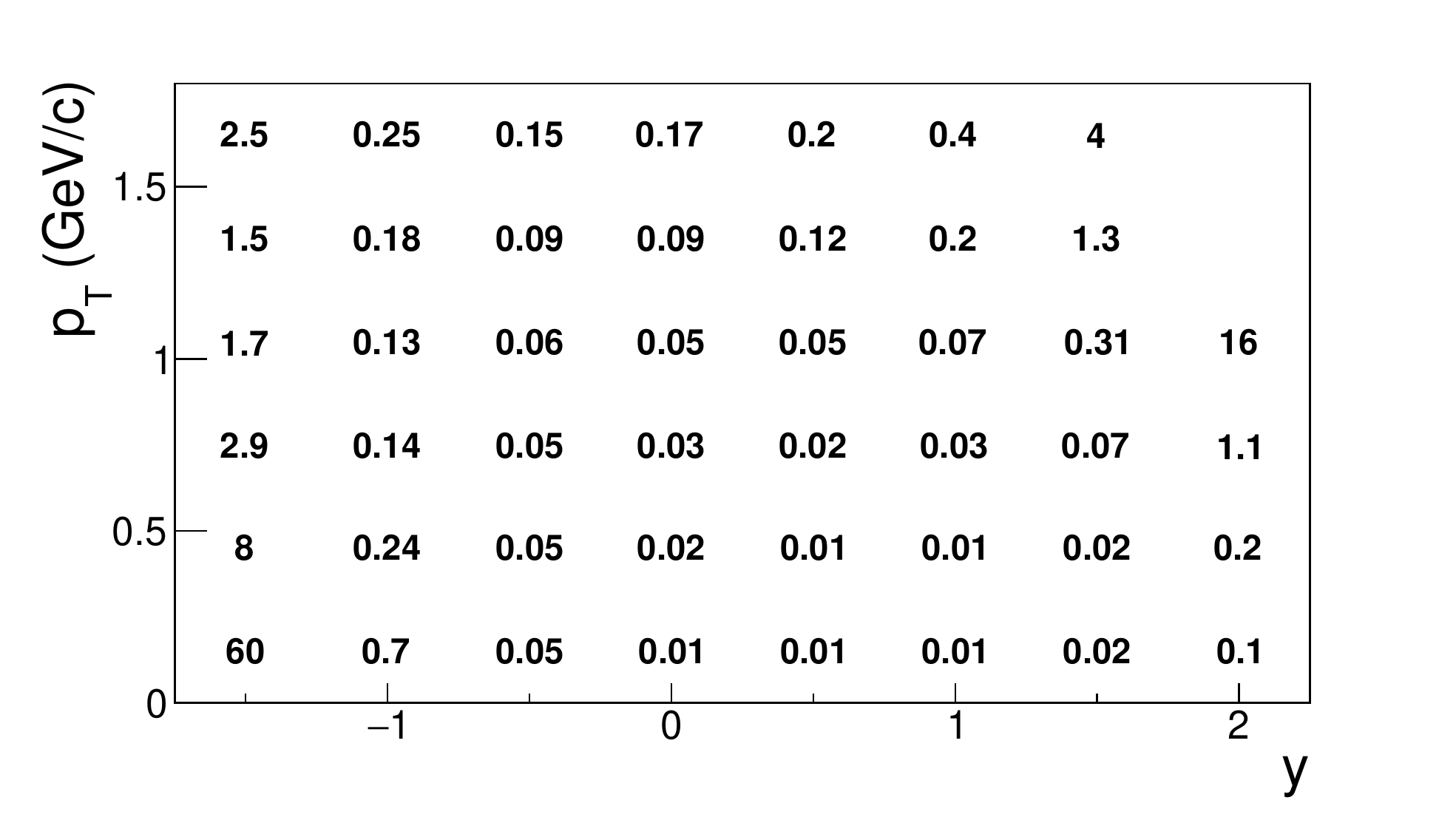} 
  \includegraphics[width=0.49\textwidth]{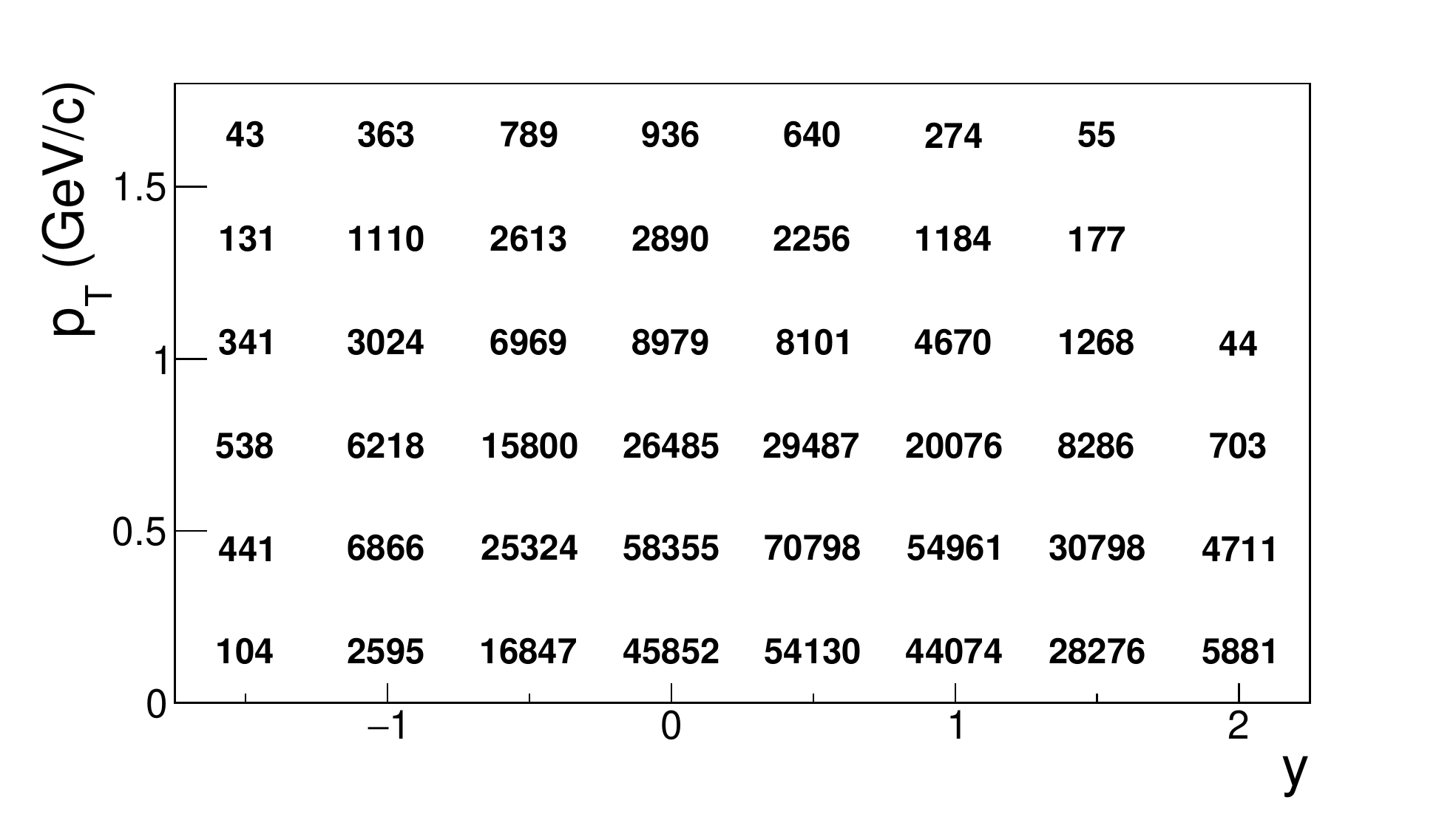}
  \includegraphics[width=0.49\textwidth]{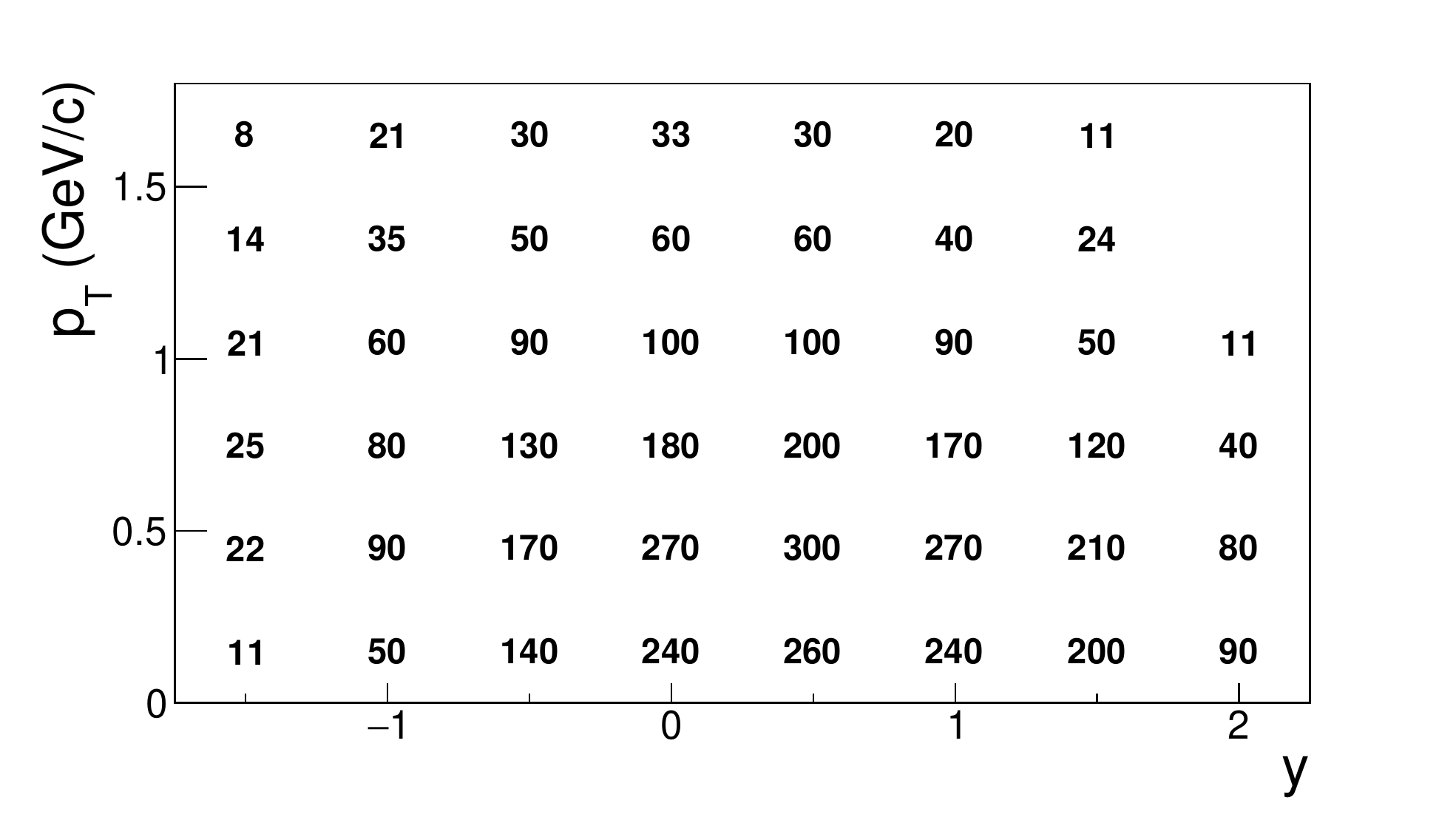}
  \caption[]{\textit{Top}: Monte-Carlo correction factors (see Eq.~\ref{eq:cmc}) in each ($y, p_T$) bin (\textit{left}) and corresponding statistical uncertainties (\textit{right}). \textit{Bottom}: uncorrected bin-by-bin multiplicities of $K^{0}_{S}$ (\textit{left}) and statistical uncertainties (\textit{right}).}
  \label{fig:Cmc}
\end{figure*}

Uncorrected bin-by-bin $K^{0}_{S}$ multiplicities and their statistical uncertainties are shown in Fig.~\ref{fig:Cmc}~(\textit{bottom}).

\subsection{Correction factors}
\label{sec:Correction_factors}

A detailed Monte Carlo simulation was performed to compute the correction for losses due to the trigger bias, geometrical acceptance, reconstruction efficiency, as well as the selection cuts applied in the analysis. The correction factors are based on $9.5 \times 10^7$ inelastic \textit{p+p} events at 158~\GeVc produced by the \EposLong event 
generator~\cite{Werner:2005jf, Pierog:2009zt}. Particles in the generated events were tracked through the \NASixtyOne apparatus using the \GeantThree package~\cite{GEANT}. The TPC response was simulated by dedicated software packages which take into account all known detector effects. The simulated events were reconstructed with the same software as used for real events and the same selection cuts were applied and \dedx identification was replaced by matching of simulated and reconstructed tracks. The branching ratio of $K^{0}_{S}$ decays are taken into account in the \GeantThree software package. For each $y$ and $p_T$ bin, the correction factor $c_{MC}(y,p_T)$ was calculated as:
	
\begin{equation}
	c_{MC} (y,p_T) =\left.\frac{n_{MC}^{gen}(y,p_T)}{N_{MC}^{gen}} \right/  \frac{n_{MC}^{acc}(y,p_T)}{N_{MC}^{acc}}~,
\label{eq:cmc}	
\end{equation}
where: 
	\begin{itemize}
		\item [-] $n_{MC}^{gen}(y,p_T)$ is the number of $K^{0}_{S}$ generated in a given ($y, p_T$) bin,
		\item [-] $n_{MC}^{acc}(y,p_T)$ is the number of  reconstructed $K^{0}_{S}$ in a given ($y, p_T$) bin. To derive this numbers the invariant mass distribution of the reconstructed $\pi^{+}$ and $\pi^{-}$ track pairs that pass all selection requirements was formed. The number of reconstructed $K^{0}_{S}$ is then obtained by following the same extraction procedure as for real data, described in Sec.~\ref{s:signal_extraction}.
		\item [-] $N_{MC}^{gen}$ is the number of generated inelastic \textit{p+p} interactions ($9.5 \times 10^7$),
		\item [-] $N_{MC}^{acc}$ is the number of accepted \textit{p+p} events ($5.4 \times 10^7$). 
	\end{itemize}

The loss of the $K^{0}_{S}$ mesons due to the \dedx cut is corrected with an additional factor:
\begin{equation}
	c_{dE/dx} = \frac{1}{\epsilon ^2} = 1.005~,
\end{equation}
where $\epsilon = 0.9973$ is the probability for the pions to be detected within $\pm 3\sigma$ around the nominal Bethe-Bloch value.

The double-differential yield of $K^{0}_{S}$ per inelastic event in bins of ($y, p_T$) is calculated as follows:

\begin{equation}
    \frac{d^2 n}{dy\, dp_T} (y, p_T) = \frac{c_{dE/dx} \cdot c_{MC}(y,p_T)}{\Delta y \, \Delta p_T} \cdot \frac{n_{K^{0}_{S}}(y,p_T)}{N_{events}}~,
\label{eq:dndydpt}
\end{equation}

where: 
\begin{itemize}
	\item [-] $ c_{dE/dx}$, $c_{MC}(y,p_T)$ are the correction factors described above,
	\item [-] $\Delta y$ and $\Delta p_T$ are the bin widths,
	\item [-] $n_{K^{0}_{S}}(y,p_T)$ is the uncorrected number of $K^{0}_{S}$, obtained by the signal extraction procedure described in Sec.~\ref{s:signal_extraction},
	\item [-] $N_{events}$ is the number of events after cuts.
\end{itemize}


\subsection{Statistical uncertainties}
\label{s:statistical_uncertainties}

The statistical uncertainties of the corrected double-differential yields (see Eq.~\ref{eq:dndydpt}) receive contributions from the statistical uncertainty of the correction factors $c_{MC}(y,p_T)$ and the statistical uncertainty  of the uncorrected number of $K^{0}_{S}$ ($\Delta N_{K^{0}_{S}} (y, p_T)$). The statistical uncertainty of the former receives two contributions, the first, $\alpha$, caused by the loss of inelastic interactions due to the event selection and the second, $\beta$, connected with the loss of $K^{0}_{S}$ candidates due to the $V^0$ selection:
\begin{equation}
	c_{MC} (y,p_T) =\left.\frac{n_{MC}^{gen}(y,p_T)}{N_{MC}^{gen}} \right/  \frac{n_{MC}^{acc}(y,p_T)}{N_{MC}^{acc}} = \left.\frac{N_{MC}^{acc}}{N_{MC}^{gen}} \right/  \frac{n_{MC}^{acc}(y,p_T)}{n_{MC}^{gen}(y,p_T)} = \frac{\alpha}{\beta(y,p_T)}~,
\end{equation}

The error of $\alpha$ is calculated assuming a binomial distribution:

\begin{equation}
	\Delta \alpha = \sqrt{\frac{\alpha(1-\alpha)}{N_{MC}^{gen}}}~,
\end{equation}

The error of $\beta$ is calculated according to formula:

\begin{equation}
\Delta \beta (y,p_T) = \sqrt{\left(\frac{\Delta n_{MC}^{acc}(y,p_T)}{n_{MC}^{gen}(y,p_T)} \right)^2+\left(\frac{n_{MC}^{acc}(y,p_T) \cdot \Delta n_{MC}^{gen}(y,p_T)}{(n_{MC}^{gen}(y,p_T))^2} \right)^2}~,
\end{equation}

where $\Delta n_{MC}^{acc}(y,p_T)$ is the uncertainty of the fit, and $\Delta n_{MC}^{gen}(y,p_T)= \sqrt{n_{MC}^{gen}(y,p_T)}$. The equation for $\Delta c_{MC} (y,p_T)$ can be written as:

\begin{equation}
    \Delta c_{MC} (y,p_T) = \sqrt{\left(\frac{\Delta \alpha}{\beta}\right)^2+\left(-\frac{\alpha \cdot \Delta \beta}{\beta^2}\right)^2}~,
\end{equation}

The statistical uncertainties $\Delta n_{K^{0}_{S}} (y, p_T)$ of the corrected number of $K^{0}_{S}$ are:

\begin{equation}
	\Delta \frac{d^2 n}{dy dp_T} ( y, p_T ) =\sqrt{\left ( \frac{c_{dE/dx} \cdot c_{MC}(y,p_T)}{N_{events}\, \Delta y \, \Delta p_T} \right )^2 \Delta n_{K^{0}_{S}}^2 (y, p_T) + \left ( \frac{c_{dE/dx} \cdot n_{K^{0}_{S}}(y,p_T)}{N_{events}\, \Delta y \, \Delta p_T} \right)^2 \Delta c_{MC}^2(y, p_T)}~. 
\end{equation}

\subsection{Systematic uncertainties}
\label{s:systematic_uncertainties}

Three possible sources of the systematic uncertainties related to event selection criteria, the track and $V^0$ selection criteria and the signal extraction procedure, are included.

The following effects were considered in the calculation of the systematic uncertainties:

\begin{itemize}
	\item [(i)] The uncertainties related to event selection criteria were estimated by performing the analysis with the following changes:
		\begin{itemize}
			\item 
			Simulations were done with and without the S4 trigger condition. One half of the difference between these two results was taken as a contribution to the systematic uncertainty, which is 3-10\%.
			\item Vertex \coordinate{z} position was changed from -590 < \coordinate{z}~(\cm) < -572 to -588 < \coordinate{z}~(\cm) < -574. The uncertainty due to variation of the selection window was estimated to be up to 2\%.
		\end{itemize}	
	
	\item [(ii)] The uncertainties related to track and $V^0$ selection criteria were estimated by performing the ana\-lysis with the following changes compared to the original values:
		\begin{itemize}
			\item the minimum required number of clusters in both VTPCs for $K^{0}_{S}$ decay products was changed from 15 to 10 and 20 yielding a possible bias up to 4\%,
			\item the standard \dedx cut used for identification of $K^{0}_{S}$ decay products was changed from $\pm 3\sigma$ to $\pm 2.5\sigma$ and $\pm 3.5\sigma$ from the nominal Bethe-Bloch value yielding a possible bias up to 5\%,
			\item DCA cut for daughter tracks at the $V^0$ decay vertex was changed from 1~\cm to 0.5~\cm and 1.5~\cm~yielding a possible bias up to 4\%, 
			\item the impact parameter cut for the daughters tracks was varied by 50\%: $\left(\frac{b_x}{2}\right)^2+(b_y)^2<0.125$ and $\left(\frac{b_x}{2}\right)^2+(b_y)^2<0.375$ yielding a possible bias up to 2\%, 
			\item the $\Delta$\coordinate{z} cut was changed from |$\Delta$\coordinate{z}| < $e^{3.1+0.42 \cdot y}$ to |$\Delta$\coordinate{z}| < $e^{2.96+0.47 \cdot y}$ and |$\Delta$\coordinate{z}| < $e^{3.24+0.38 \cdot y}$ yielding a possible bias up to 3\%,
			\item the $cos\Phi$ cut was varied with respect to the nominal values yielding a possible bias up to 3\%. The range of cut values is listed in 
			Table~\ref{tab:tab_CosPhi}.
		\end{itemize}

	\item [(iii)] The uncertainty due to the signal extraction procedure was estimated by:
		\begin{itemize}
			\item changing the background fit function from a $2^{nd}$ order to a $3^{rd}$ order polynomial yielding a possible bias up to 4\%,
			\item changing the invariant mass range over which the uncorrected number of $K^{0}_{S}$ was integrated from $m_0\pm 3\Gamma$ to $\pm 2.5\Gamma$ and $\pm 3.5\Gamma$ yielding a possible bias up to 2\%, 
			\item calculating the uncorrected number of $K^{0}_{S}$ as the sum of entries after background fit subtraction instead of the integral of the Lorentzian signal function yielding a possible bias up to 4\%,
			\item changing the region of the fit from [0.35-0.7]~\GeVcc to [0.4-0.65]~\GeVcc~yielding a possible bias up to 3\%.
		\end{itemize}

\end{itemize}

\begin{table} [ht]
\small
	\centering
	\begin{tabular}{|c|c|c|c|c|}
		\hline
		\multicolumn{5}{|c|}{Maximal $|cos\Phi|$ allowed} \\
		\hline
		$y_{min}$ & $y_{max}$ & original & new lower & new upper \\
		\hline
		-0.25 & 0.25 & 0.95 & 0.925 & 0.975 \\
		\hline
		0.25 & 0.75 & 0.9 & 0.85 & 0.95 \\
		\hline
		0.75 & 1.25 & 0.8 & 0.75 & 0.85 \\
		\hline
		1.25 & 1.75 & 0.5 & 0.4 & 0.6 \\
		\hline
	\end{tabular}
\caption{Numerical values for $cos\Phi$ cut used for systematic uncertainties calculation.}
\label{tab:tab_CosPhi}
\end{table}

The maximum deviations are determined for every group of possible sources, which contribute to the systematic uncertainty. The systematic uncertainty was calculated as the square root of the sum of squares of the described possible biases assuming that they are uncorrelated. 
This procedure was used to estimate systematic uncertainties of all final quantities presented in this paper - yield in each ($y,p_T$) bin, inverse slope parameter of transverse momentum spectrum, yield in each rapidity bin and mean multiplicity.  

\subsection{Mean lifetime measurements}

\begin{figure*}[hbt!]
    \centering
	\includegraphics[width=0.7\textwidth]{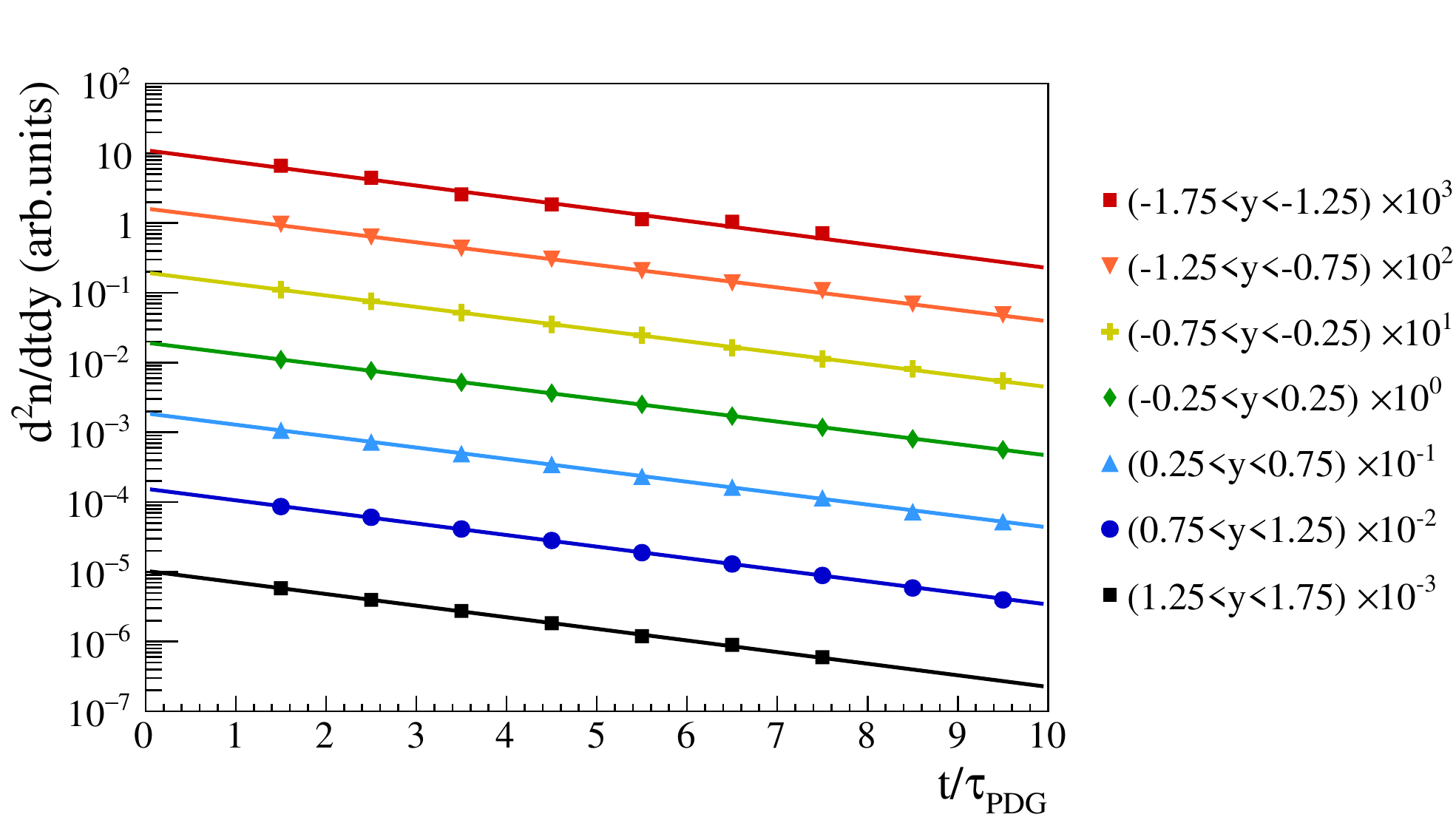}
	\caption[]{(Color online) Corrected lifetime distribution for $K^{0}_{S}$ mesons produced in inelastic \textit{p+p} interaction at 158~\GeVc. The curves show the result of the exponential fit function used to obtain the mean lifetime. Statistical uncertainties are smaller than marker size and not visible on the plot. 
	}
	\label{fig:dndtau}
\end{figure*}

\begin{figure*}[hbt!]
    \centering
	\includegraphics[width=0.7\textwidth]{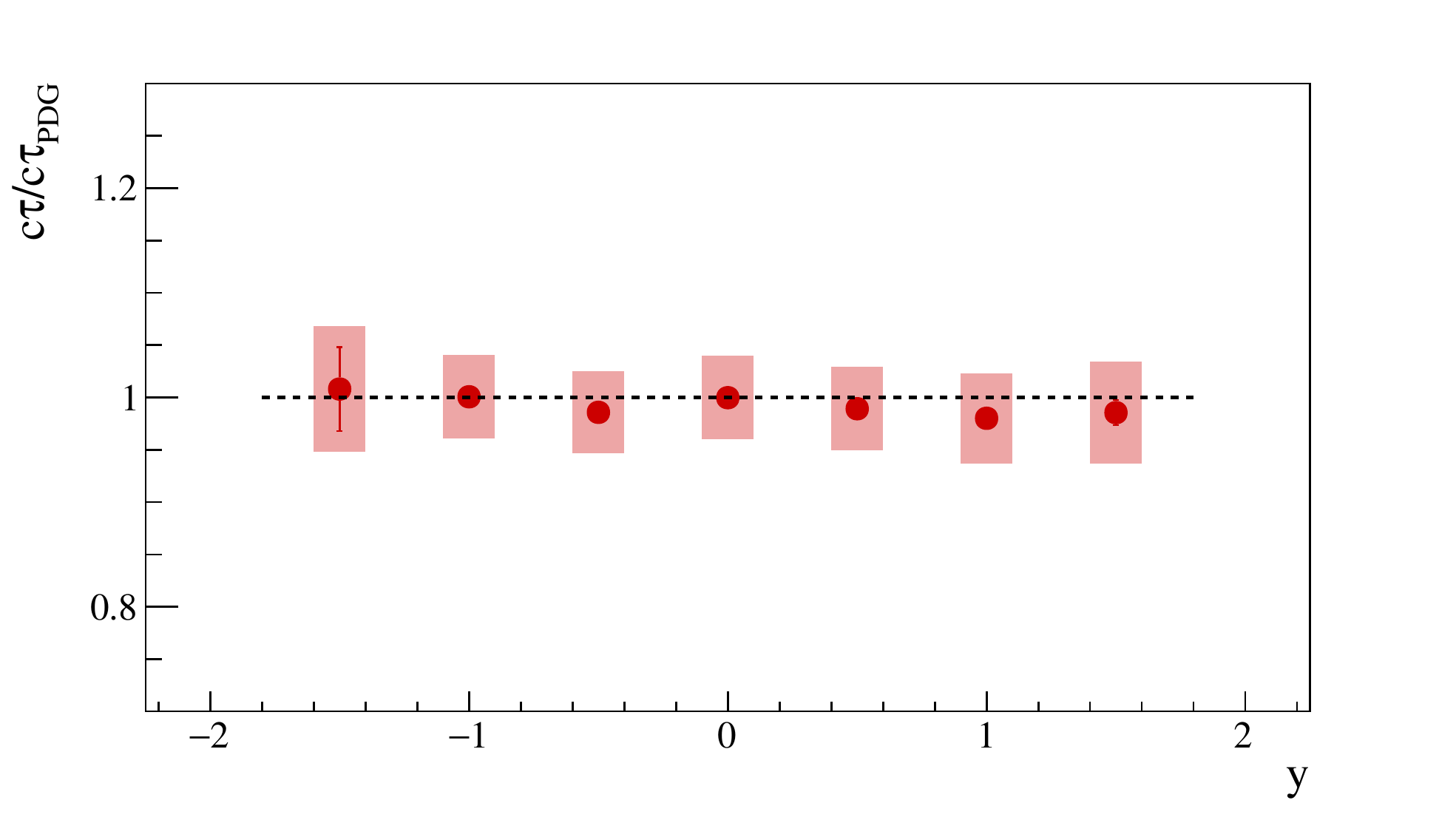}
	\caption[]{(Color online) Mean lifetime obtained from fits to the lifetime distributions of Fig.~\ref{fig:dndtau} with statistical (vertical bars) and systematic (shaded boxes) uncertainties versus the rapidity $y$.
    }
	\label{fig:lifetime}
\end{figure*}

The reliability of the $K^0_S$ reconstruction and of the correction procedure can be validated by studying the lifetime distribution of the analysed $K^0_S$. The lifetime of each $K^0_S$ candidate was calculated from the $V^0$ path length and its velocity. The lifetime distributions corrected for experimental biases (Sec.~\ref{sec:Correction_factors}) in all 7 rapidity bins were fitted by an exponential distribution to obtain proper lifetimes (see Fig.~\ref{fig:dndtau}). The obtained ratio of the measured mean lifetime to the PDG~\cite{PDG} value $c\tau_{PDG}=2.6844$~\cm is shown in Fig.~\ref{fig:lifetime} as a function of rapidity. The measured $K^{0}_{S}$ lifetime agrees within uncertainties with the PDG value and thus confirms the quality of the analysis. 

\section{Results}\label{sec:results}

This section presents the new \NASixtyOne results on inclusive $K^{0}_{S}$ meson spectra in inelastic \textit{p+p} interactions at beam momentum 158~\GeVc. The spectra refer to weakly decaying $K^{0}_{S}$ mesons produced in strong interaction processes and are corrected for experimental biases and the branching ratio.


\subsection{Transverse momentum spectra}

\begin{figure*}
	\centering
	\includegraphics[width=0.6\textwidth]{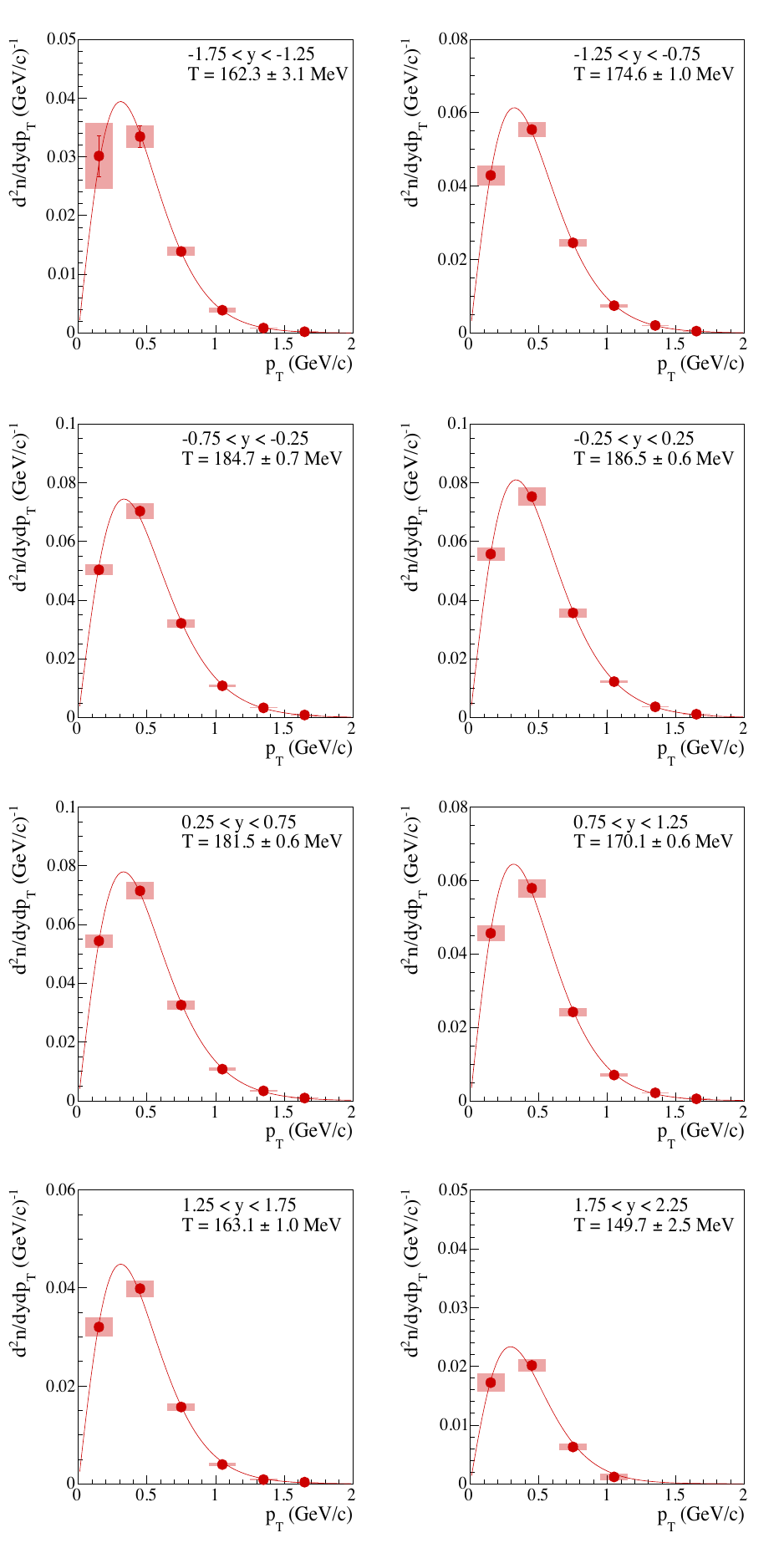}
\vspace{-0.2cm}
	\caption[]{(Color online) Double-differential $K^{0}_{S}$ spectra in inelastic \textit{p+p} interaction at 158~\GeVc in bins of ($y, p_T$) as obtained from Eq.~\ref{eq:dndydpt}. Measured points are shown as red full circles. The solid red curve is obtained from a fit to the data points using the exponential function Eq.~\ref{eq:fit_to_dndydpt}. Statistical uncertainties are indicated by vertical bars (for some points smaller than the symbol size). Red shaded boxes show systematic uncertainties. Only statistical uncertainties are used in the fits as they are uncorrelated bin-to-bin. The numerical values of the data points are listed in Table~\ref{tab:dndydpt}.
}
	\label{fig:sys_hpt}
\end{figure*}

Double differential $K^{0}_{S}$ yields listed in Table~\ref{tab:dndydpt} represent the main result of this paper. Yields are determined in 8 consecutive rapidity bins in the interval $-1.75<y<2.25$ and 6 transverse momentum bins in the interval $0.0<p_T~(\GeVc)<1.8$. The transverse momentum distributions are shown in Fig.~\ref{fig:sys_hpt}.

\begin{table}[hbt!]
    \centering
	\small
	\begin{tabular}{|c|c|c|c|c|}
	\hline
	& \multicolumn{4}{c|}{$y$} \\
	\cline{2-5}
	$p_T$ (\GeVc) & (-1.75;-1.25) & (-1.25;-0.75) & (-0.75;-0.25) & (-0.25;0.25) \\
	\cline{1-5}
	(0.0;0.3) & 30.1 $\pm$ 3.5 $\pm$ 5.6 & 43 $\pm$ 1 $\pm$ 3 & 50.3 $\pm$ 0.5 $\pm$ 1.9 & 55.7 $\pm$ 0.3 $\pm$ 2.2 \\
	\cline{1-5}
	(0.3;0.6) & 33 $\pm$ 2 $\pm$ 2 & 55.4 $\pm$ 0.8 $\pm$ 2.1 & 70.3 $\pm$ 0.6 $\pm$ 2.7 & 75.2 $\pm$ 0.4 $\pm$ 3.1 \\
	\cline{1-5}
	(0.6;0.9) & 13.9 $\pm$ 0.8 $\pm$ 0.8 & 24.6 $\pm$ 0.4 $\pm$ 1.1 & 32.1 $\pm$ 0.3 $\pm$ 1.5 & 35.6 $\pm$ 0.3 $\pm$ 1.5 \\
	\cline{1-5}
	(0.9;1.2) & 3.9 $\pm$ 0.3 $\pm$ 0.4 & 7.4 $\pm$ 0.2 $\pm$ 0.5 & 10.8 $\pm$ 0.2 $\pm$ 0.5 & 12.2 $\pm$ 0.2 $\pm$ 0.5 \\
	\cline{1-5}
	(1.2;1.5) & 0.8 $\pm$ 0.1 $\pm$ 0.2 & 2.07 $\pm$ 0.08 $\pm$ 0.15 & 3.32 $\pm$ 0.09 $\pm$ 0.15 & 3.7 $\pm$ 0.1 $\pm$ 0.2\\
	\cline{1-5}
	(1.5;1.8) & 0.22 $\pm$ 0.05 $\pm$ 0.07 & 0.52 $\pm$ 0.04 $\pm$ 0.04 & 0.89 $\pm$ 0.04 $\pm$ 0.04 & 1.15 $\pm$ 0.05 $\pm$ 0.05\\
	\hline
	& \multicolumn{4}{c|}{$y$} \\
	\cline{2-5}
	$p_T$ (\GeVc) & (0.25;0.75) &  (0.75;1.25) & (1.25;1.75) & (1.75;2.25)\\
	\cline{1-5}
	(0.0;0.3) & 54.4 $\pm$ 0.3 $\pm$ 2.3 & 45.6 $\pm$ 0.3 $\pm$ 2.1 & 32.0 $\pm$ 0.3 $\pm$ 2.0 & 17.2 $\pm$ 0.3 $\pm$ 1.5 \\
	\cline{1-5}
	(0.3;0.6) & 71.4 $\pm$ 0.4 $\pm$ 3.0 & 57.9 $\pm$ 0.3 $\pm$ 2.5 & 39.9 $\pm$ 0.3 $\pm$ 1.7 & 20.1 $\pm$ 0.4 $\pm$ 1.0 \\
	\cline{1-5}
	(0.6;0.9) & 32.6 $\pm$ 0.3 $\pm$ 1.5 & 24.2 $\pm$ 0.2 $\pm$ 1.1 & 15.7 $\pm$ 0.3 $\pm$ 0.8 & 6.3 $\pm$ 0.4 $\pm$ 0.5 \\
	\cline{1-5}
	(0.9;1.2) & 10.7 $\pm$ 0.2 $\pm$ 0.5 & 7.0 $\pm$ 0.2 $\pm$ 0.4 & 3.9 $\pm$ 0.2 $\pm$ 0.4 & 1.2 $\pm$ 0.3 $\pm$ 0.5 \\
	\cline{1-5}
	(1.2;1.5) & 3.34 $\pm$ 0.10 $\pm$ 0.19 & 2.14 $\pm$ 0.09 $\pm$ 0.14 & 0.86 $\pm$ 0.13 $\pm$ 0.15 &\\
	\cline{1-4}
	(1.5;1.8) & 0.91 $\pm$ 0.05 $\pm$ 0.09 & 0.51 $\pm$ 0.05 $\pm$ 0.07 & 0.34 $\pm$ 0.09 $\pm$ 0.10 & \\
	\cline{1-5}
	\hline
\end{tabular}

\caption{Double differential $K^{0}_{S}$ yields in bins of ($y, p_T$). The first uncertainty is statistical, while the second one is systematic.
}
\label{tab:dndydpt}
\end{table}


The transverse momentum spectra can be described by an exponential function:

\begin{equation}
f(p_T) = A \cdot p_T \cdot \exp \left(\frac{\sqrt{p_{T}^{2}+m_{0}^{2}}}{T}\right)~,
\label{eq:fit_to_dndydpt}
\end{equation}

where $m_{0}$ is the mass of the $K^{0}_{S}$ and $T$ is the inverse slope parameter. Fits of Eq.~\ref{eq:fit_to_dndydpt} to the data points provide the values of $T$ in each rapidity bin which are listed in Table~\ref{tab:dndy} and in the legend of the panels in Fig.~\ref{fig:sys_hpt}.


\subsection{Rapidity distribution and mean multiplicity}
\label{sec:mean_multip}

\begin{figure*}
	\centering
	\includegraphics[width=0.7\textwidth]{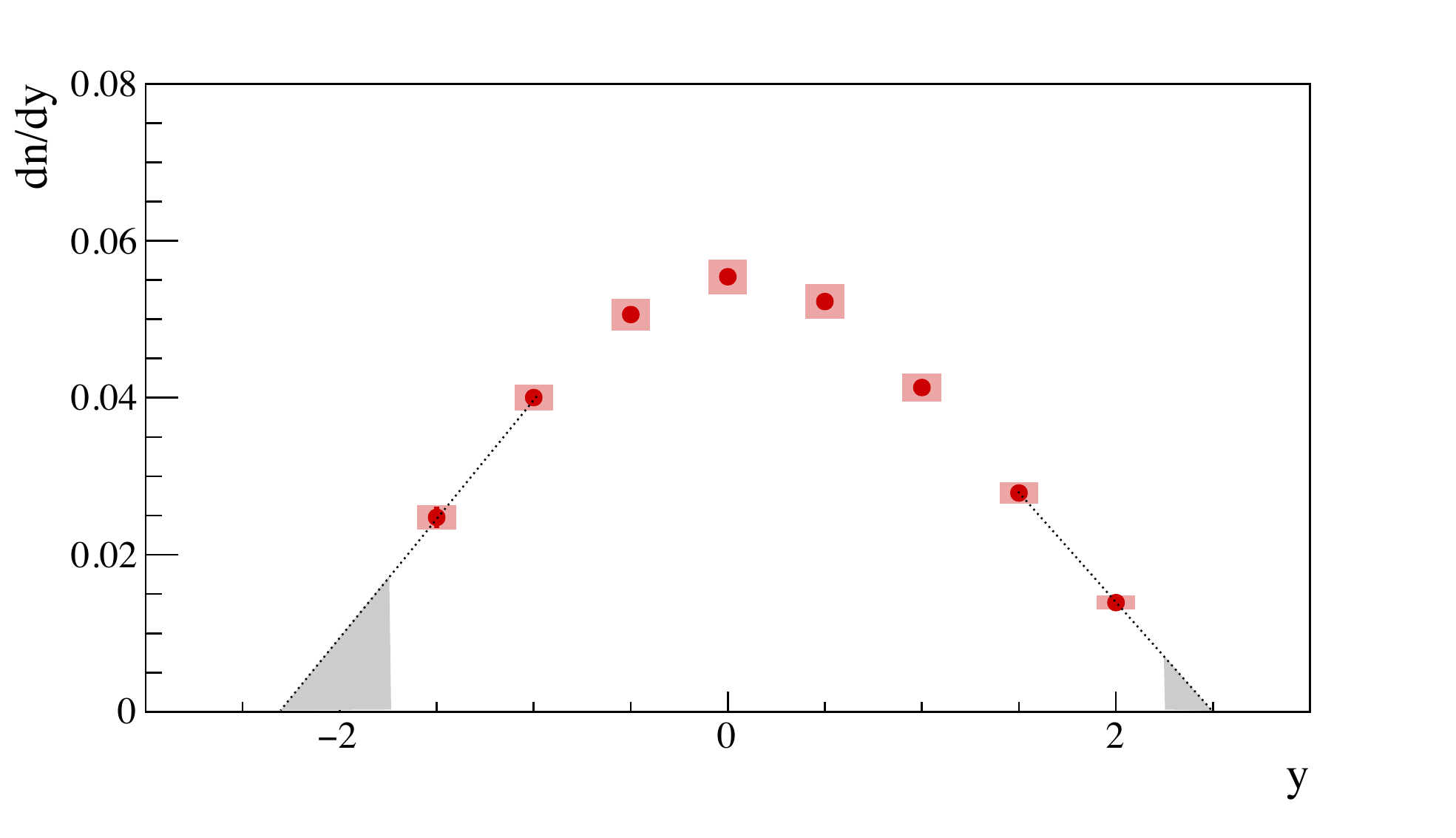}
	\caption[]{(Color online) Rapidity distribution $dn/dy$ obtained by \pt-integration. Statistical uncertainties are shown by vertical bars (often smaller than the marker size), while a red shaded boxes indicates systematic uncertainties. The black dotted lines show the connection line between the last two points on both sides, and the grey areas are the contributions of the extrapolation to the mean multiplicity of $K^{0}_{S}$ mesons. The numerical data are listed in Table~\ref{tab:dndy}.}
	\label{fig:dndy}
\end{figure*}

Kaon yields in each rapidity bin were obtained from the corresponding measured transverse momentum distributions. The small fraction of $K^{0}_{S}$ at high \pt outside of the acceptance was determined using Eq.~\ref{eq:fit_to_dndydpt}.
The resulting $\frac{dn}{dy}$ spectrum of $K^{0}_{S}$ mesons produced in inelastic \textit{p+p} interactions at 158~\GeVc is plotted in Fig.~\ref{fig:dndy}. 

\begin{table}
\small
\centering
\begin{tabular}{|c|c|c|c|}
	\hline
	$y$ & T (MeV) & $\frac{dn}{dy} \times 10^{-3}$ & extrapolated fraction (\%) \\
	\hline
    (-1.75;-1.25) & 162.3 $\pm$ 3.1 $\pm$ 7.0 & 24.8 $\pm$ 1.2 $\pm$ 1.5 & 0.21 \\
	\hline
	(-1.25;-0.75) & 174.6 $\pm$ 1.0 $\pm$ 2.3 & 40.0 $\pm$ 0.4 $\pm$ 1.6 & 0.35 \\
	\hline
	(-0.75;-0.25) & 184.7 $\pm$ 0.7 $\pm$ 0.7 & 50.6 $\pm$ 0.3 $\pm$ 1.8 & 0.56 \\
	\hline
	(-0.25;0.25) & 186.5 $\pm$ 0.6 $\pm$ 0.8 & 55.4 $\pm$ 0.2 $\pm$ 2.1 & 0.58 \\
	\hline
	(0.25;0.75) & 181.5 $\pm$ 0.6 $\pm$ 0.8 & 52.3 $\pm$ 0.2 $\pm$ 2.1 & 0.47 \\
	\hline
	(0.75;1.25) & 170.1 $\pm$ 0.6 $\pm$ 1.6 & 41.3 $\pm$ 0.2 $\pm$ 1.8 & 0.31 \\
	\hline
	(1.25;1.75) & 163.1 $\pm$ 1.0 $\pm$ 2.0 & 27.9 $\pm$ 0.2 $\pm$ 1.3 & 0.23 \\
	\hline
	(1.75;2.25) & 149.7 $\pm$ 2.5 $\pm$ 4.2 & 13.9 $\pm$ 0.2 $\pm$ 0.8 & 3.33 \\
	\hline
	\end{tabular}
	\vspace{0.5 cm}
	\caption{First column shows the rapidity range. In the second column the values of the inverse slope parameter are listed with its statistical and systematic uncertainties. The third column shows the numerical values of the \pt-integrated yields presented in Fig.~\ref{fig:dndy} with statistical and systematic uncertainties. In the last column contribution to the dn/dy in \% of the extrapolation to the unmeasured transverse momentum region.}

\label{tab:dndy}
\end{table}

The mean multiplicity of $K^{0}_{S}$ mesons was calculated as the sum of measured points in Fig.~\ref{fig:dndy} and the integral below linear functions through the last two measured points on both sides representative for the unmeasured region (for rapidity $y<-1.75$ and $y>2.25$). The statistical uncertainty of $\langle K^{0}_{S} \rangle$ was calculated as the square root of the sum of the squares of the statistical uncertainties of the contributing bins. The systematic uncertainty was calculated as the square root of squares of systematic uncertainty described in Sec.~\ref{s:systematic_uncertainties} and half of the extrapolated yield. To estimate the systematic uncertainty of the method used to determine the mean multiplicity of $K^{0}_{S}$, the rapidity distribution was also fitted using a single Gaussian or two Gaussians symmetrically displaced from midrapidity. The deviations of the results
of these fit from $\langle K^{0}_{S} \rangle$ is included as a contribution to the final systematic uncertainty. 


The numerical value of the total yield of $\langle K^{0}_{S} \rangle$ is:
\begin{equation}
    \langle K^{0}_{S} \rangle = 0.162 \pm 0.001 (stat.) \pm 0.011 (sys.)
\end{equation} 


\section{Comparison with published world data and model predictions}\label{sec:comparison}


This section compares the new \NASixtyOne measurement of $K^0_S$ production in inelastic \textit{p+p} interactions at 158~\GeVc with publicly available world data as well as with predictions from microscopic and statistical models \EposLong~\cite{Werner:2005jf, Pierog:2009zt}, UrQMD~3.4~\cite{Bass:1998ca, Bleicher:1999xi}, SMASH~2.0~\cite{Mohs:2019iee} and PHSD~\cite{Cassing:2008sv, Cassing:2009vt}.

\begin{figure*}[ht]
	\centering
	\includegraphics[width=0.7\textwidth]{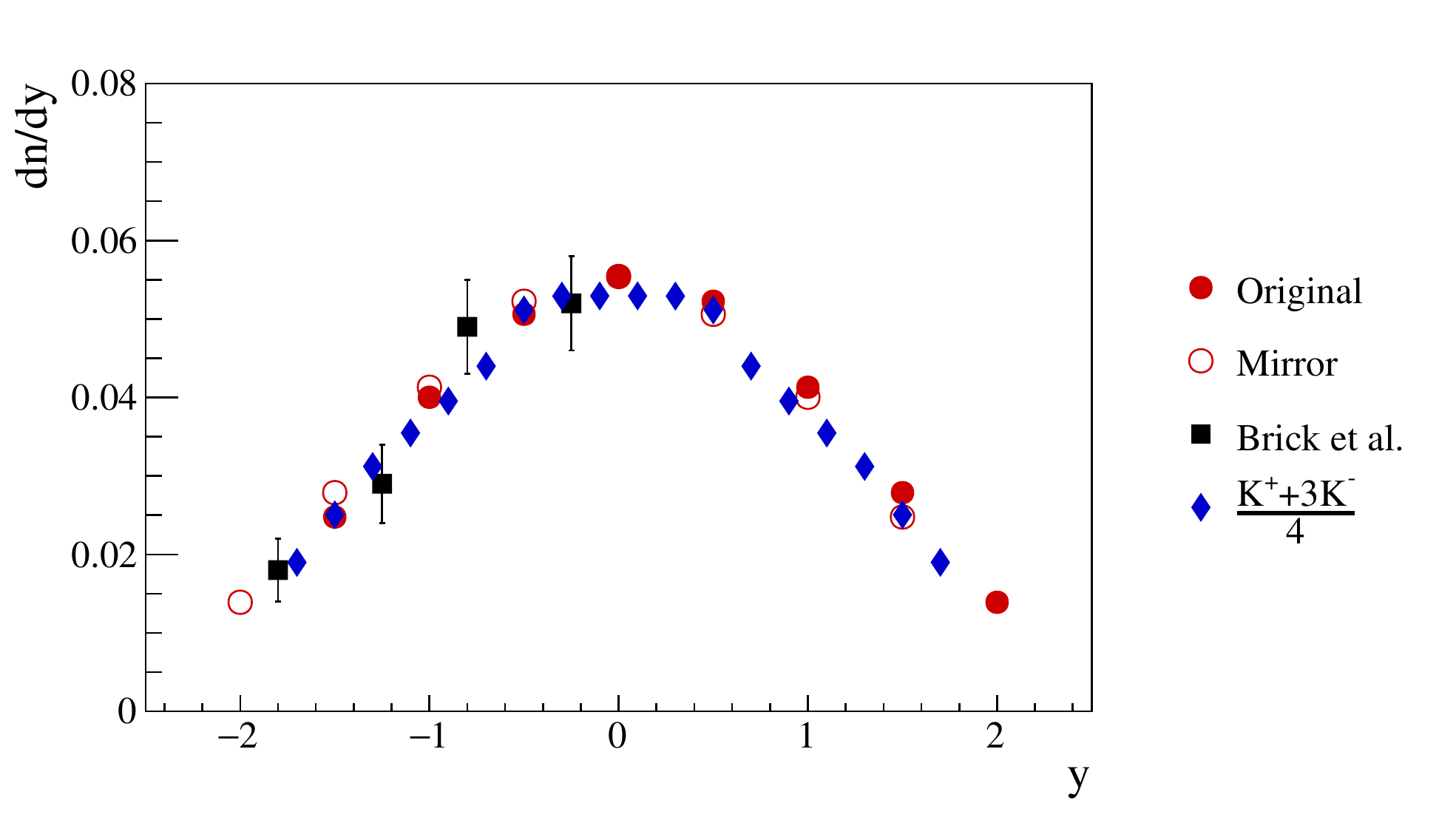}
	\caption[]{(Color online) Rapidity distribution $dn/dy$ of $K^{0}_{S}$ mesons in inelastic \textit{p+p} interactions at 158~\GeVc. Solid red-coloured circles correspond to the \NASixtyOne results (systematic uncertainties not shown on the plot), open circles are mirrored values, black squares represent results from Brick $et$ $al.$ at FNAL~\cite{Brick:1980vj} and blue full diamonds show results obtained from the formula $\frac{1}{4}(N_{K^+}+3 \cdot N_{K^-})$ using charged kaon yields recently measured by \NASixtyOne at the same beam momentum~\cite{NA61SHINE:2017fne}. }
	\label{fig:dndy_comparison}
\end{figure*}

The $K^{0}_{S}$ rapidity spectrum from \NASixtyOne is compared in Fig.~\ref{fig:dndy_comparison} to the results from Brick $et$ $al.$ at 147 \GeVc~\cite{Brick:1980vj} as well as with predictions obtained from $K^+$ and $K^-$ yields published by \NASixtyOne for inelastic \textit{p+p} interactions at 158~\GeVc~\cite{NA61SHINE:2017fne}. These predictions are based on valence-quark counting arguments~\cite{Doble:1994fb} and lead to the formula  $\frac{1}{4}(N_{K^+}+3 \cdot N_{K^-})$. Such a model was applied earlier for p+C~\cite{Abgrall:2015hmv} and for p+Be~\cite{Bonesini:2001iz} interactions. The measured $K^0_S$ yields are seen to agree with this prediction and with the measurement of Ref.~\cite{Brick:1980vj} within statistical errors. 

\begin{figure*}[ht]
	\centering
	\vspace{0.5 cm}
	\includegraphics[width=0.7\textwidth]{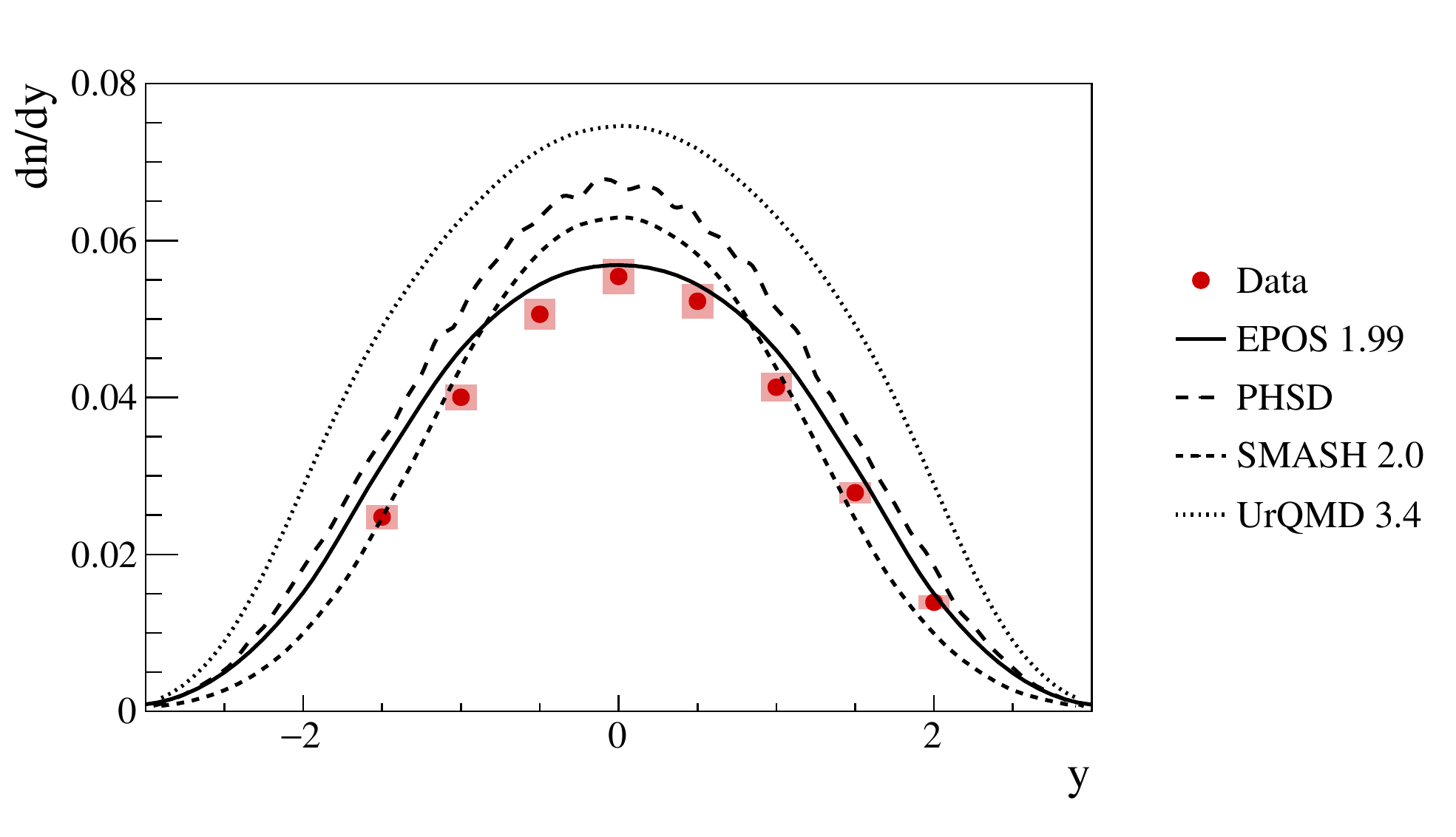}
	\caption[]{(Color online) Comparison of the results on the $K^0_S$ rapidity distribution with predictions of theoretical models. Coloured red circles show the new measurements of \NASixtyOne. The black curves show predictions of models: \EposLong (solid), PHSD (long dashed), SMASH~2.0 (short dashed) and UrQMD~3.4 (dotted). 
	}
	\label{fig:models}
\end{figure*}

Figure~\ref{fig:models} presents a comparison of the \NASixtyOne measurements with predictions of the \EposLong, 
PHSD, SMASH~2.0 and UrQMD~3.4 models. Only \EposLong describes the experimental data fairly well. All other models overpredict the $K^0_S$ yield by $10 - 20\%$. The shape of the rapidity distribution is also reproduced by the PHSD model. 

\begin{figure*}[hbt!]
	\centering
	\includegraphics[width=0.7\textwidth]{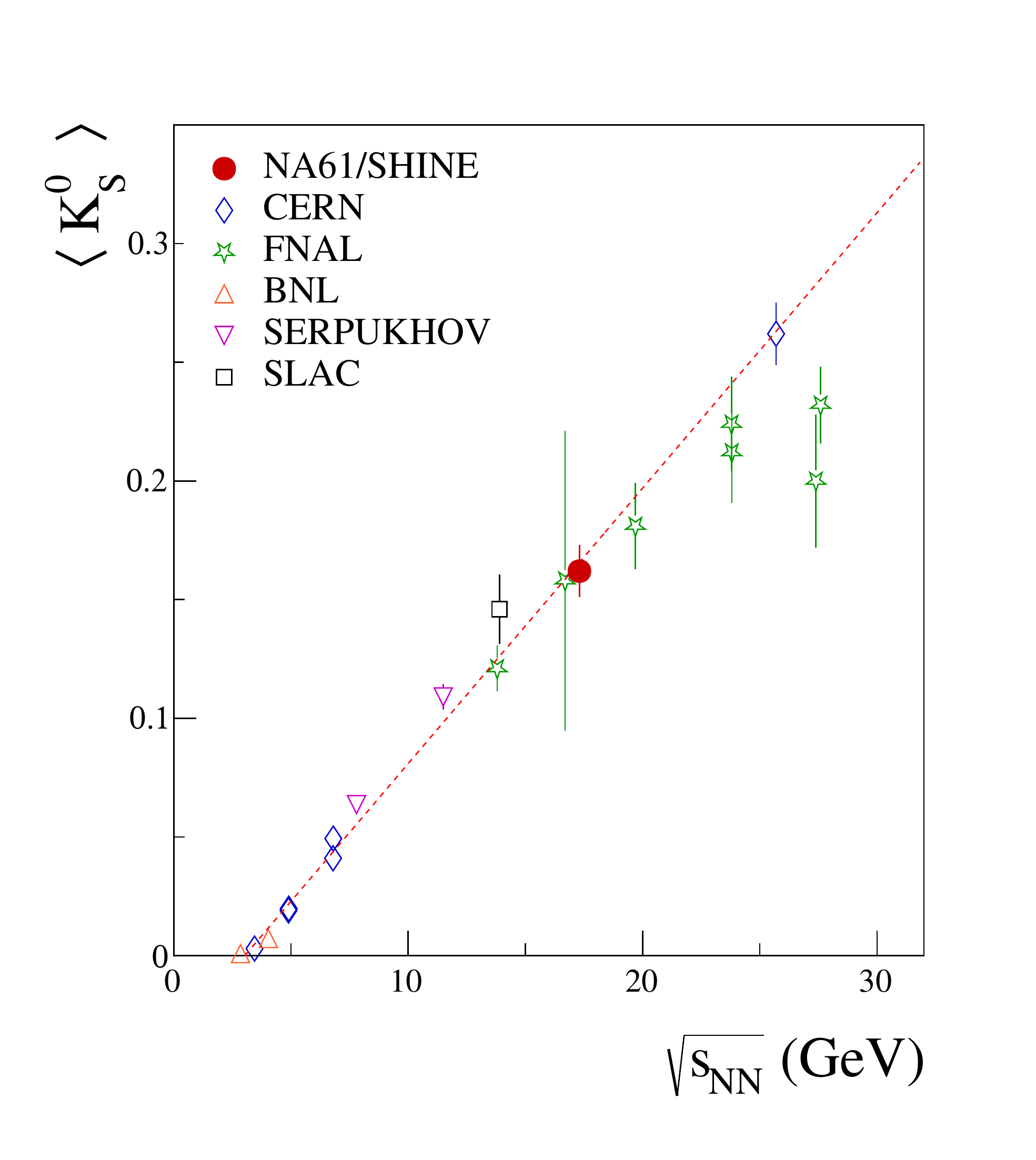}
	\vspace{-0.5 cm}
	\caption[]{(Color online) Collision energy dependence of mean multiplicity of $K^{0}_{S}$ mesons produced in \textit{p+p} interactions. The solid coloured red circle shows the measurement by \NASixtyOne  presented with its systematic uncertainty. The results published by other experiments are shown by open diamonds coloured in blue (CERN), open stars coloured in green (FNAL), open triangle pointing up coloured in orange (BNL), open triangle pointing down coloured in magenta (SERPUKHOV) and open square coloured in black (SLAC). All the data points are shown with combined statistical and systematic uncertainties. The references for the plotted data points are given in the text.
    }
	\label{fig:multiplicity_comparison}
\end{figure*}

The mean multiplicity of $K^{0}_{S}$ mesons in \textit{p+p} collisions measured by \NASixtyOne at $\sqrt{s_{NN}}=17.3$~\GeV is compared in Fig.~\ref{fig:multiplicity_comparison} with the world data in the range from 3 - 32~\GeV ~\cite{Louttit:1961zz, Alexander:1967zz, Firebaugh:1968rq, Blobel:1973jc, Fesefeldt:1979, Bogolyubsky:1988ei, Ammosov1976, AstonGarnjost:1975im, Chapman:1973fn, Brick:1980vj, Jaeger1974pk, Sheng1976, Lopinto:1980ct, Bailly1987, Kass:1979nf, Kichimi:1979te}. The measured values are seen to rise linearly with collision energy $\sqrt{s_{NN}}$.

\vspace{-0.3 cm}

\newpage
\section{Summary}\label{sec:summary}

This paper presents the new \NASixtyOne measurement of $K^{0}_{S}$ meson production via its $\pi^{+} \pi^{-}$ decay mode in inelastic \textit{p+p} collisions at beam momentum 158~\GeVc ($\sqrt{s_{NN}}=17.3$~\GeV). Spectra of transverse momentum (up to 1.8 \GeVc), as well as a distribution of rapidity (from -1.75 to 2.25), are presented. The mean multiplicity, obtained from the \pt-integrated and extrapolated rapidity distribution, is $(0.162 \pm 0.001 \pm 0.011)$, where the first uncertainty is statistical and the second systematic. The rapidity distribution is in agreement with results from other experiments at nearby beam momenta. 
Mean multiplicity from model calculations deviate by up to $20\%$ from the measurements. The \EposLong model provides the best predictions for the experimental data. 
The results of $K^{0}_{S}$ production in proton-proton interactions presented in this paper 
significantly improve, with their high statistical precision, the knowledge of strangeness production in elementary interactions.


\section*{Acknowledgements}
We would like to thank the CERN EP, BE, HSE and EN Departments for the
strong support of NA61/SHINE.

This work was supported by
the Hungarian Scientific Research Fund (grant NKFIH 123842\slash123959),
the Polish Ministry of Science
and Higher Education (grants 667\slash N-CERN\slash2010\slash0,
NN\,202\,48\,4339 and NN\,202\,23\,1837), the National Science Centre Poland (grants
2014\slash14\slash E\slash ST2\slash00018,
2014\slash15\slash B\slash ST2\slash\- 02537 and
2015\slash18\slash M\slash ST2\slash00125,
2015\slash 19\slash N\slash ST2\slash01689,
2016\slash23\slash B\slash ST2\slash00692,
DIR\slash WK\slash\- 2016\slash 2017\slash\- 10-1,
2017\slash\- 25\slash N\slash\- ST2\slash\- 02575,
2018\slash30\slash A\slash ST2\slash00226,
2018\slash31\slash G\slash ST2\slash03910,
2019\slash34\slash H\slash ST2\slash00585,
2016\slash21\slash D\slash ST2\slash01983),
WUT ID-UB, 
the Russian Science Foundation, grant 16-12-10176 and 17-72-20045,
the Russian Academy of Science and the
Russian Foundation for Basic Research (grants 08-02-00018, 09-02-00664
and 12-02-91503-CERN),
the Russian Foundation for Basic Research (RFBR) funding within the research project no. 18-02-40086,
the Ministry of Science and Higher Education of the Russian Federation, Project "Fundamental properties of elementary particles and cosmology" No 0723-2020-0041,
the European Union's Horizon 2020 research and innovation programme under grant agreement No. 871072,
the Ministry of Education, Culture, Sports,
Science and Tech\-no\-lo\-gy, Japan, Grant-in-Aid for Sci\-en\-ti\-fic
Research (grants 18071005, 19034011, 19740162, 20740160 and 20039012),
the German Research Foundation DFG (grants GA\,1480/8-1 and project 426579465), the Bulgarian Ministry of Education and Science within the National Roadmap
for Research Infrastructures 2020-2027, contract No. D01-374/18.12.2020, Ministry of Education and Science of the Republic of Serbia (grant OI171002), Swiss
Nationalfonds Foundation (grant 200020\-117913/1), ETH Research Grant
TH-01\,07-3 and the Fermi National Accelerator Laboratory (Fermilab), a U.S. Department of Energy, Office of Science, HEP User Facility managed by Fermi Research Alliance, LLC (FRA), acting under Contract No. DE-AC02-07CH11359 and the IN2P3-CNRS (France).

\bibliographystyle{include/na61Utphys}
\bibliography{main.bbl}

\newpage
{\Large The \NASixtyOne Collaboration}
\bigskip
\begin{sloppypar}

\noindent
A.~Acharya$^{\,10}$,
H.~Adhikary$^{\,10}$,
K.K.~Allison$^{\,26}$,
N.~Amin$^{\,5}$,
E.V.~Andronov$^{\,22}$,
T.~Anti\'ci\'c$^{\,3}$,
V.~Babkin$^{\,20}$,
M.~Baszczyk$^{\,14}$,
S.~Bhosale$^{\,11}$,
A.~Blondel$^{\,4}$
% ...

\end{sloppypar}

\noindent
% ...
$^{3}$~Ru{\dj}er Bo\v{s}kovi\'c Institute, Zagreb, Croatia\\
$^{4}$~LPNHE, University of Paris VI and VII, Paris, France\\
$^{5}$~Karlsruhe Institute of Technology, Karlsruhe, Germany\\
% ...
$^{10}$~Jan Kochanowski University in Kielce, Poland\\
$^{11}$~Institute of Nuclear Physics, Polish Academy of Sciences, Cracow, Poland\\
% ...
$^{14}$~AGH - University of Science and Technology, Cracow, Poland\\
% ...
$^{20}$~Joint Institute for Nuclear Research, Dubna, Russia\\
% ...
$^{22}$~St. Petersburg State University, St. Petersburg, Russia\\
% ...
$^{26}$~University of Colorado, Boulder, USA\\
% ...

\end{document}